\documentclass[conference]{IEEEtran}

\usepackage{algorithm}
\usepackage[noend]{algpseudocode} 
\usepackage{algorithmicx}

\usepackage{amsmath,amssymb,amsfonts}
\usepackage{graphicx}
\usepackage{textcomp}
\usepackage{xcolor}
\usepackage{colortbl}
\usepackage{subfigure}
\usepackage{optidef} 
\usepackage{multicol}
\usepackage{multirow}
\usepackage{hyperref}
\usepackage{listings}
\usepackage{lipsum} 
\usepackage[normalem]{ulem}
\usepackage[shortlabels]{enumitem}

\definecolor{todo}{RGB}{190, 110, 190}
\newcommand{\todo}[1]{{\color{todo}#1}}
\newcommand{\done}[1]{{\color{blue}#1}}
\definecolor{sdr}{rgb}{0.0, 0.65, 0.31}

\begin{document}

\title{\textit{Infer-EDGE}: Dynamic DNN Inference Optimization in `Just-in-time' Edge-AI  Implementations
}


\author{
Motahare Mounesan$^\ast$, Xiaojie Zhang$^\$$, Saptarshi Debroy$^\ast$\\
$^\ast$City University of New York, $^\$$Hunan First Normal University\\ Emails: \textit{mmounesan@gradcenter.cuny.edu, xiaojie.zhang@hnfnu.edu.cn, saptarshi.debroy@hunter.cuny.edu}}




\maketitle


\begin{abstract}
Balancing mutually diverging performance metrics, such as, end-to-end latency, accuracy, and device energy consumption is a challenging undertaking for deep neural network (DNN) inference  in `Just-in-Time' edge environments that are inherent resource constrained and loosely coupled. 
In this paper, we design and develop {\em Infer-EDGE} framework that seeks to strike such balance for latency-sensitive video processing applications. First, using comprehensive benchmarking experiments, we develop intuitions about the trade-off characteristics, which is then used by the framework to develop an Advantage Actor-Critic (A2C) Reinforcement Learning (RL) approach that can choose optimal run-time DNN inference parameters 
aligns the performance metrics based on the application requirements. 
Using real world DNNs and a hardware testbed, we evaluate the benefits of {\em Infer-EDGE} framework in terms of device energy savings, inference accuracy improvement, and end-to-end inference latency reduction.  \\
\end{abstract}



\vspace{-0.1in}




\section{Introduction}
Artificial intelligence (AI), specifically, deep neural networks (DNN) play a critical role in 
mission critical applications, e.g., public safety, tactical scenarios, search and rescue, and emergency triage, 
that are often edge native. However, unlike traditional edge which is often deployed as part of network infrastructure, `just-in-time' edge environments are ad-hoc deployments, with or without connectivity to the cloud, setup to support mission-critical use cases that extensively host pre-trained AI workloads (mostly DNN inference) with strict latency and accuracy requirements. 


However, implementing such `just-in-time' Edge-AI comes with challenges as 
executing computationally expensive DNN models locally on resource-constrained IoT devices (i.e., `local-only' processing) result in significant resource consumption and rapid battery drainage. However, the alternative option of offloading the entire computation to the edge server (i.e., `remote-only' processing) is not always viable for three primary reasons: 1) unlike traditional service provider operated edge, `just-in-time' edge servers might have limited computational capacity, 2) for privacy-preserving purposes, many mission critical applications prohibit sending unprocessed information (e.g., raw images) which can violate application/data privacy/security requirements, and 3) low/critically-low upload data rates in such environments 
often make sending such large data streams at high-quality significantly time consuming and result in high energy drainage of IoT devices~\cite{9985008}.

Thus, there is a need to optimize DNN inference for `just-in-time' edge-AI implementations. One approach to achieve such
involves mitigating the computational cost by adjusting a given DNN model's architecture~\cite{teerapittayanon2016branchynet,9151283}. However, such model modifications, in many cases, can impact {\em model inference accuracy}, which in turn affects the application performance.
The alternative approach embraces runtime (i.e., during model inference) optimization strategies, often referred to as partial offloading, 
where a segment of the DNN model is processed on the IoT device until the data is sufficiently reduced, making the transmission cost  to the remote server more manageable \cite{9985008}.
Such partial offloading, also referred to as DNN partitioning/splitting, can be affected by unique characteristics of the DNN, such as, number of convolutional layers, computational complexity, and output data size at each layer, among other things~\cite{10.1145/3093337.3037698, 10.1145/3527155}.
Therefore, such optimization techniques can impact the {\em end-to-end inference latency}, as well as the {\em energy consumption} of IoT devices for data transmission to remote servers. 
This fundamental three-way trade-off between {\em end-to-end inference latency}, {\em model inference accuracy}, and {\em IoT device energy consumption} metrics (henceforth referred to as `latency-accuracy-energy') while performing DNN inference optimization at `just-in-time' edge is non-trivial to achieve and is grossly under-explored in the current literature. 



Additionally, 
the conventional wisdom for addressing such multi-objective  optimization problems in distributed computing involves classic stochastic optimization techniques (e.g., Lyapunov optimization), which often require a comprehensive understanding of the behavior of system parameters~\cite{9039590, zhang2023effect}. However, this might become challenging for `just-in-time' edge environments due to: i) the inherent dynamic nature of such environments in terms of resource availability and application requirements, ii) limited availability of continuous in-system measurements, and iii) underlying optimization problem's high dimensionality.
Instead, reinforcement learning (RL) approaches have proven effective to address such challenges. By learning through the interactions with the environment, RL based techniques offer a viable alternative in navigating the complexities of such scenarios, where traditional optimization techniques may face limitations.



In this paper, we propose {\em Infer-EDGE}, a framework for `just-in-time' Edge-AI aiming to strike a balance between 
`latency-accuracy-energy' running heterogeneous DNNs and under varying edge resource availability.
The framework allows IoT devices to cache different model versions with varying latency-accuracy-energy profiles which provides the edge system to choose an execution profile, which is a combination of selecting an optimized version of a given DNN model among many such pre-cached version and selecting a partition cut point layer for the chosen version for collaborative inference with the edge server. 
Such execution profile selection is framed as an optimization problem, striving to maximize the long-term average of a customizable performance function. 
The proposed function flexibly incorporates `latency-accuracy-energy' metrics to satisfy specific requirements of different `just-in-time' edge-AI implementation scenarios. 
The optimization problem, reframed as a Markov Decision Process (MDP), is solved using an Advantage Actor-Critic (A2C) based RL algorithm that learns about the environment through actions and rewards and converges at an optimal trade-off point that optimizes the `latency-accuracy-energy' metrics. Overall, the main contributions of this paper are as follows:

\vspace{-0.05in}
\begin{enumerate}[leftmargin=*,itemsep=0pt]
    \item We perform a comprehensive benchmark study with real world DNNs and a `just-in-time' edge testbed to demonstrate the latency-accuracy-energy trade-off and develop system optimization intuitions.    
    
    
    
    \item The proposed framework is customizable that dictates two design choices: the selection of an optimized version of a given DNN model and the selection of cut point layer for the chosen version. More specifically, the framework can switch between lightweight and heavyweight versions, cached at the IoT devices and the edge server, with each version and corresponding cut layer offering unique latency-accuracy-energy profile. 
    
    \item 
    The underlying A2C model seamlessly incorporates inputs from the device's battery status, activity profile, available bandwidth, and dispatched tasks to continuously learn the system dynamics through an iterative set of actions and rewards, without requiring prior knowledge about the system characteristics.
    
    \item We validate the stability of the proposed A2C algorithm and substantiate the framework with object classification DNNs and the testbed, illustrating how system parameters, such as battery level, transmission rate, device activity level, and DNN model affect decision-making regarding the execution plan. 
    The results reveal that {\em Infer-EDGE} can achieve upto 77\% reduction in latency and 92\% decrease in energy consumption compared to baseline strategies, without compromising the inference accuracy.
\end{enumerate}

The rest of the paper is organized as follows. Section~\ref{sec:related_work} discusses the related work. Section~\ref{sec:motivation} demonstrates 
problem evidence analysis. Section~\ref{sec:systemmodel} introduces the solution strategy.
Section~\ref{evaluaton} discusses system evaluation. 
Section~\ref{sec:conclusion} concludes the paper.

\section{Related works and Background}
\label{sec:related_work}
\subsection{DNN inference at edge}
DNN inference on the edge mainly involves the offload of DNN tasks, which requires taking into account the distinctive characteristics of DNN. There exists certain hardware~\cite{SECthermal,jlpea12040057} and a few software strategies~\cite{8876870}. 

\subsubsection{DNN partitioning/Collaborative DNN inference}
DNN partitioning, also called collaborative DNN inference, involves dividing/split the DNN architecture into non-overlapping parts where each participant (e.g., devices, edge servers, cloud) is assigned specific parts to execute sequentially till the final result is obtained. 
The choice of the layer where the DNN is partitioned, i.e., the cut point layer depends on the system optimization objectives, and can either be static or dynamic.
For optimizing such collaborative DNN inference, striking a balance between end-to-end latency and energy consumption is considered state-of-the-art~\cite{zhang-survey}. 
One strategy is to employ a fixed head and tail model, where efforts are made to train model components for the purpose of data compression~\cite{matsubara2022bottlefit}. 
The alternative approach involves partitioning the DNN at various layers~\cite{zhang2023effect,kang2017neurosurgeon, li2019edge, laskaridis2020spinn,zhao2018deepthings,mao2017modnn}. 
\subsubsection{DNN model architecture optimization}
Broadly, collaborative DNN inference can be categorized into two approaches: solutions without DNN architecture alternation, as detailed in the previous section, and a more recent trend known as DNN model architecture optimization, such as model downsizing~\cite{10.1145/3527155,zoph2016neural}, model compression~\cite{mishra2020survey,ma2023reliability,Wiley,DNN_IIoT,dasilvabarros:hal-04497548}, and early-exit strategies~\cite{9151283, teerapittayanon2016branchynet,teerapittayanon2016branchynet,pacheco2020inference,li2019edge,huang2023elastic}, which are often motivated by DNN property (e.g., accuracy, complexity, latency) enhancement/modification. 
All such model architecture optimization techniques, essentially create modified lightweight `versions' of well-known DNNs. In essence, these versions can potentially be cached on devices and servers as real-time
execution plans, chosen based on the system's inherent dynamism.  
\textit{Unlike the existing work, our objective is to propose a more general framework wherein all such different versions of a DNN model, achieved through diverse architectural designs and compression techniques, can leverage layer-wise partitioning to facilitate latency, energy, and accuracy optimized Edge-AI deployment.}
\vspace{-0.05in}
\subsection{RL for collaborative inference}
\vspace{-0.05in}
Traditionally, distributed resource management problems for DNN inference optimization has been solved using stochastic approaches and heuristics, such as Lyapunov optimization. Works, e.g.,~\cite{9039590, zhang2023effect}, showcase problem formulation and solution strategies, such as Mixed Integer Nonlinear Programming (MINLP) and Lyapunov optimization. These approaches involve conducting thorough measurements to comprehend the characteristics of various system parameters. 
In this context, authors in \cite{zhang2021deep} employ the delayed deep deterministic policy gradient (TD3) algorithm for resource management in Industrial IoT applications. 
Authors in \cite{cui2022learning} address a similar challenge within a multi-device multi-server architecture by utilizing an LSTM-TD3 approach. In \cite{contreras2023context}, authors propose a Double Deep Q Network (DDQN) to optimize task flows in a multilayered system with heterogeneous task scheduling, with a specific focus on context switching.

Deep Reinforcement Learning (DRL)-based strategies have the potential to significantly elevate resource utilization in edge environments. Many existing schemes narrow their focus to optimize a subset of network and inference performance metrics, including communication delay, end-to-end latency, or energy consumption~\cite{10571207, 9170818, 9984691}. {\em In contrast, our work seeks to optimize DNN inference, while striking a balance among end-to-end latency, energy consumption, and accuracy. These factors can weighted according to the unique requirements of the application at hand. Our emphasis lies in utilizing DRL methods to achieve efficient resource management, ultimately delivering highly optimized inference services.} 

\begin{figure*}[h!]
    \centering
    \subfigure[]{
        \includegraphics[width=3cm]{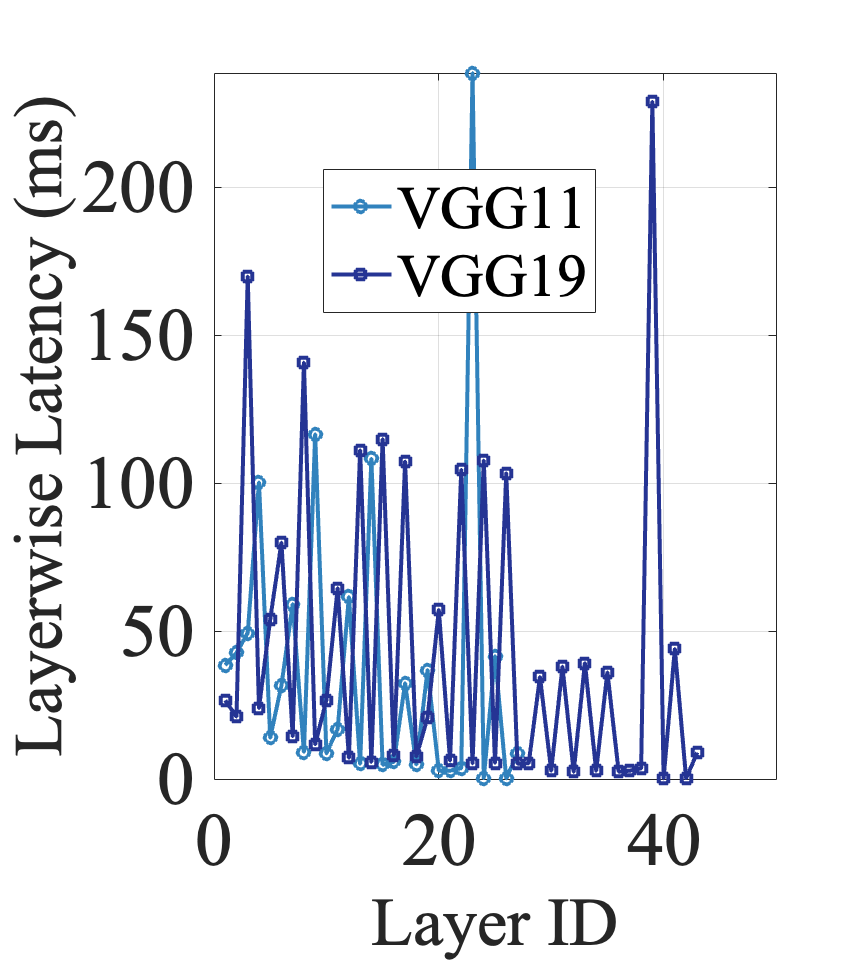}
        \label{fig:pea_vgg1}
    }\hspace{-0.16in}
    \subfigure[]{
        \includegraphics[width=2.8cm]{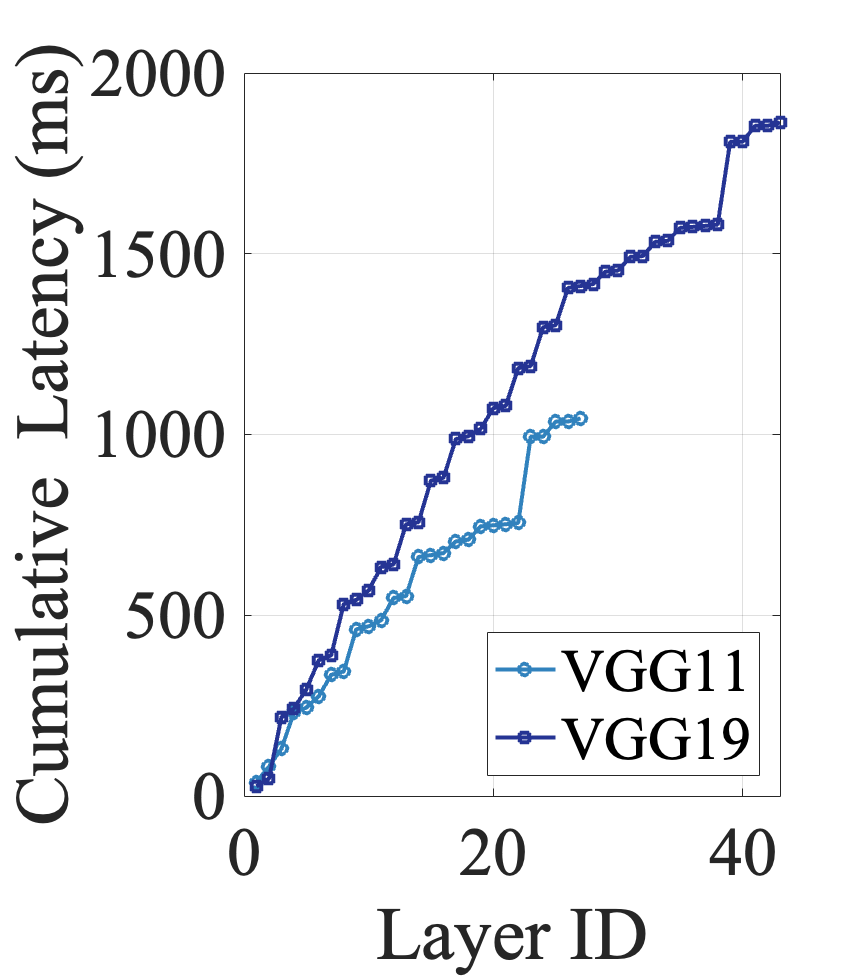}
        \label{fig:pea_vgg2}
    }\hspace{-0.16in}
    \subfigure[]{
        \includegraphics[width=2.8cm]{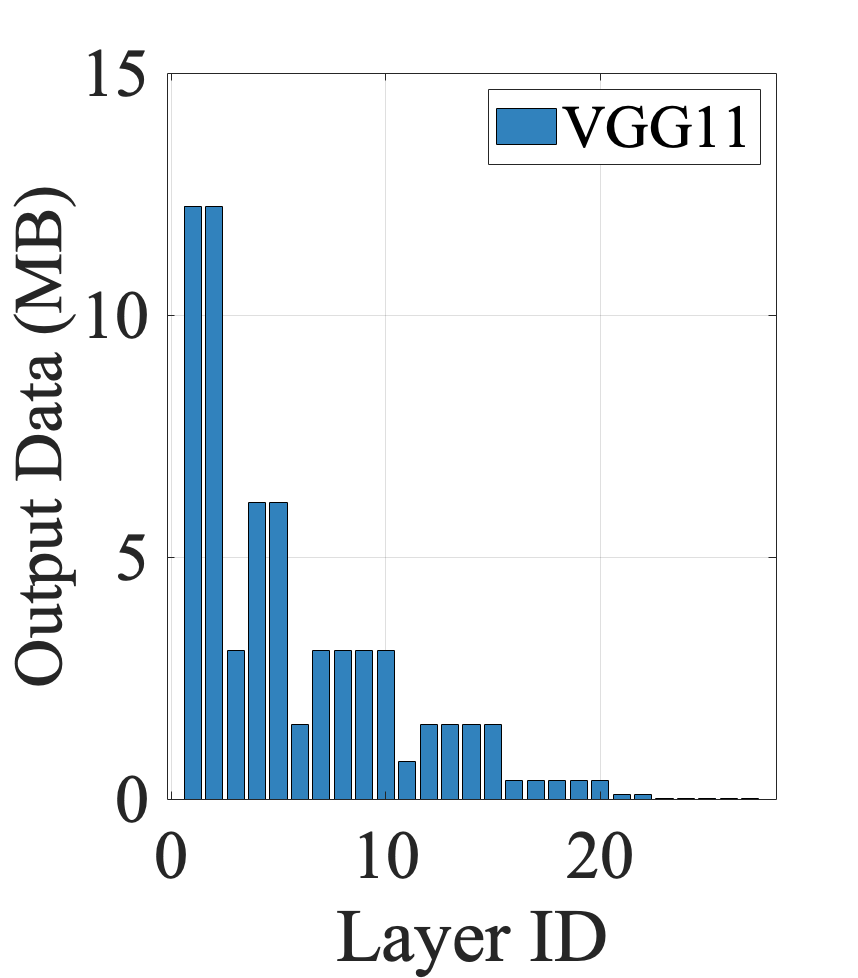}
        \label{fig:pea_vgg3}
    }\hspace{-0.16in}
    \subfigure[]{
        \includegraphics[width=2.8cm]{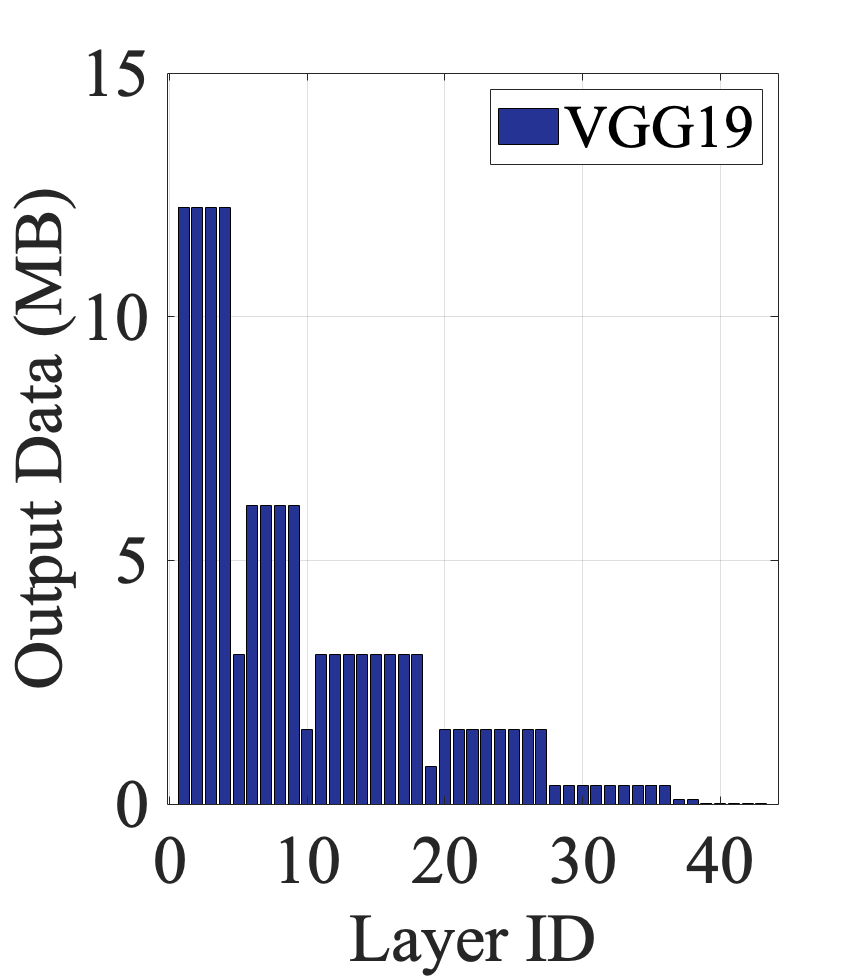}
        \label{fig:pea_vgg4}
    }\hspace{-0.16in}
    \subfigure[]{
        \includegraphics[width=2.8cm]{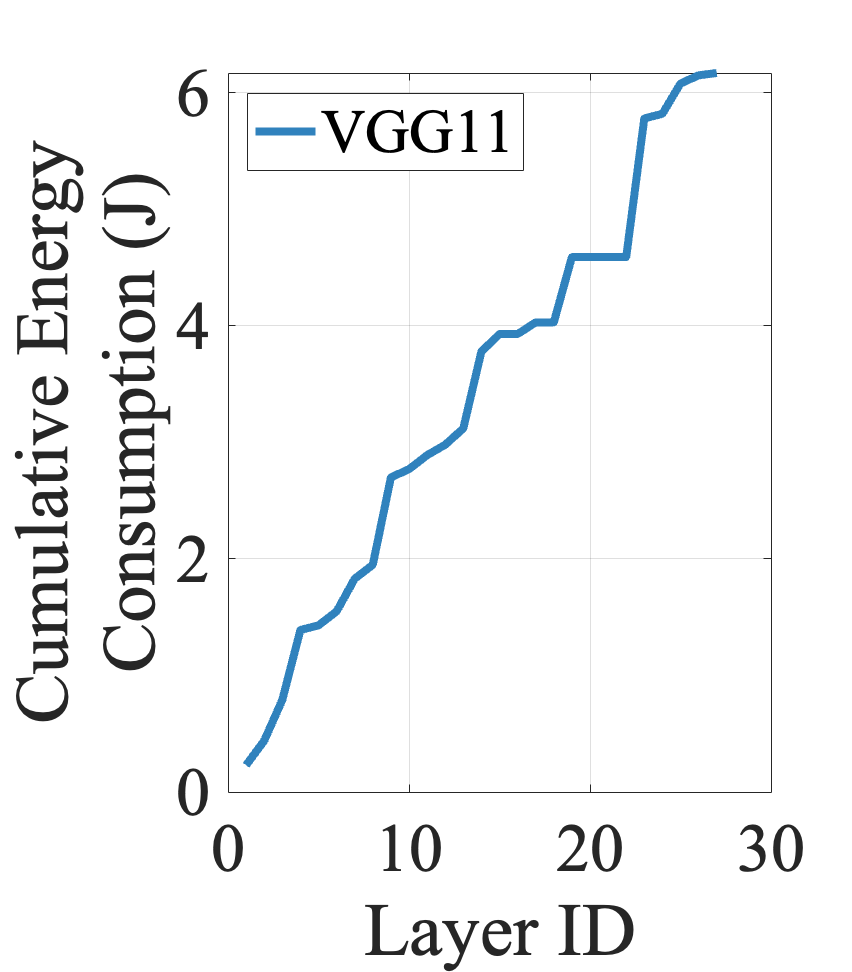}
        
        \label{fig:pea_vgg5}
    }\hspace{-0.16in}
    \subfigure[]{
        \includegraphics[width=2.8cm]{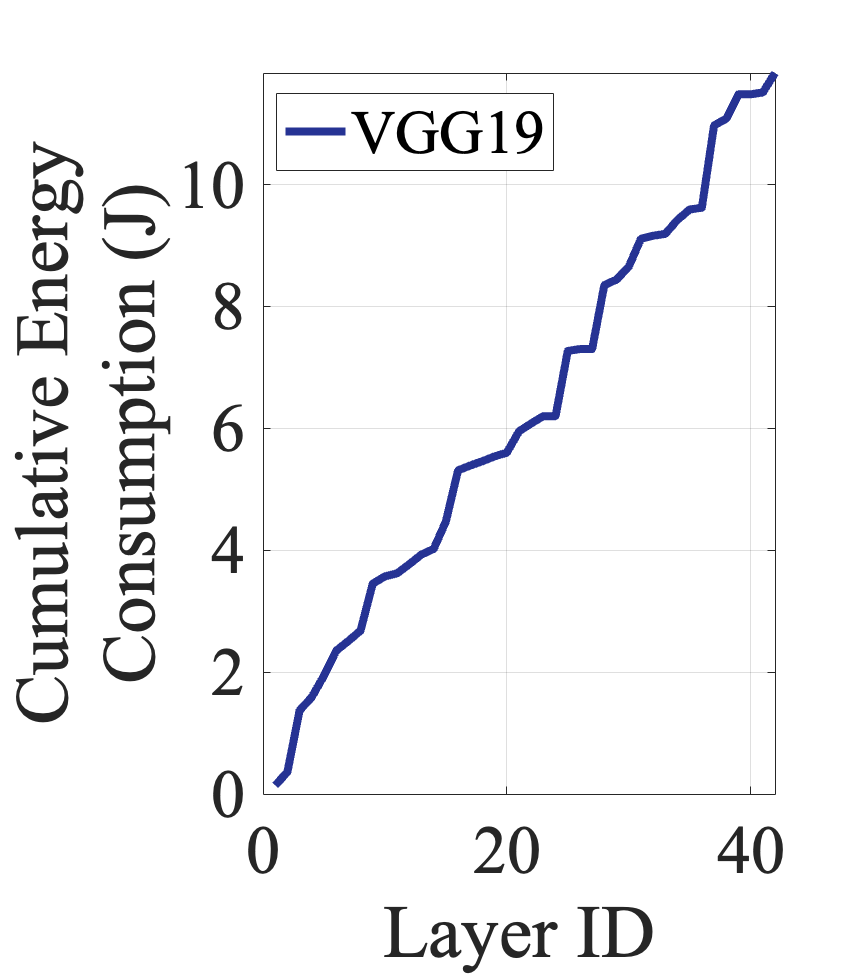}
        \label{fig:pea_vgg6}
    }
    \vspace{-0.1in}
    \caption{Layer-wise latency, output data size, and energy consumption comparisons of different versions of VGG}
    \label{fig:problemevidence}
    \vspace{-0.2in}
\end{figure*}



\section{Problem Evidence Analysis}
\label{sec:motivation}

In order to demonstrate the nature of DNN inference `latency-energy-accuracy' inter-conflict for `just-in-time' Edge-AI, we conduct benchmarking experiments on our lab testbed mimicking such environments. Specifically,
we capture the characteristics of various versions of the same DNN and layer-specific attributes within each version during DNN inference. 
For the experiments, we use image classification DNNs (viz., VGG~\cite{sengupta2019going}, ResNet~\cite{targ2016resnet}, and DenseNet~\cite{iandola2014densenet}) that are common for use cases that employ `just-in-time' edge environments. 
For the testbed, we use a IoT device-edge server pair where an NVIDIA Jetson TX2 mimics computationally capable IoT device and a Dell PowerEdge desktop with 16 cores and a CPU frequency of 3.2 GHz emulates the edge server. 
We observe the latency in processing all layers of a DNN, the corresponding output data size, and the energy consumption by IoT devices during inference to study their interplay.\\

\noindent \textbf{Model Optimization Analysis:}
As mentioned before in Section~\ref{sec:related_work}, DNN model architecture optimization  results in multiple versions of a DNN model with diverse characteristics and performance, all while targeting the same underlying model objective.
In our benchmarking experiments and throughout the remainder of the paper, we employ various versions of a specific DNN (such as VGG) as manifestations of models derived from different DNN model architecture optimization techniques. 
In this paper, we rely on PyTorch implementations of various model versions, namely VGG11, VGG19, ResNet18, ResNet50, DenseNet121, and DenseNet161. These results encompass key metrics such as `top-1' accuracy scores, inference latency, and energy consumption by Jetson TX2 (as outlined in Tab.~\ref{tab:benchmarking}). For this analysis, we present benchmarking results for different versions of VGG. However, more comprehensive results for DNN models and their versions can be found in the project GitHub repository~\cite{inferedgerepo}.

\begin{table}[t]
\scriptsize
    \centering
    \caption{Performance benchmarked on Imagenet dataset}
    \vspace{-0.1in}
    \resizebox{7cm}{!}{
    \begin{tabular}{llll}
        \hline
        Model & Accuracy (acc1) & Latency (ms) & Energy (J)\\
        \hline
        VGG11 & 69.04\% 
        & 1044.48 & 6.17\\ 
        VGG19 & 72.40\% 
         & 1862.89 & 11.83\\
        DenseNet121 & 74.43\% 
        & 4292.17 & 28.00\\
        DenseNet161 & 77.11\% 
        & 7845.49 & 50.99\\
        ResNet18 & 69.76\% 
        &  627.59 & 3.73\\
        ResNet50 & 76.15\% 
        & 984.62 & 7.46\\
        \hline
    \end{tabular}}
    \label{tab:benchmarking}
    \vspace{-0.2in}
\end{table}

As depicted in Fig.~\ref{fig:problemevidence}, we conduct layer-wise latency measurements for VGG11 and VGG19 to evaluate differences in latency attributed to the computational complexity of individual layers. These measurements are performed on our testbed using 50 images from the ImageNet validation set. Both layer-wise (Fig.~\ref{fig:pea_vgg1}) and cumulative latency (Fig.~\ref{fig:pea_vgg2}) across all layers are calculated. In Fig.~\ref{fig:pea_vgg2}, we observe that VGG19 initially performs similarly to VGG11. However, as the model progresses, the computational cost of VGG19 surpasses that of VGG11, primarily due to its larger number of layers rather than the complexity of individual layers. This finding depicts diverse layer characteristics among such models. The overarching trend emerging from these observations is the latency diversity among layers between versions.




Next, we perform experiments to observe layer-wise performance in terms of output data size, computation latency, cumulative energy consumed by TX2 (i.e., IoT device) in order to ascertain layer choices for DNN partition. It is to be noted that for this analysis, we primarily focus on the higher-level layers of the DNN models. For example, the number of convolitutional layers in DenseNet versions exceeds 100. However, due to complex dependencies within the {\em dense} blocks, we avoid partitioning the DNN in the middle of the {\em dense} block, thereby limiting our analysis to the characteristics of the higher-level layers, which sum up to 14.

When analyzing the output data size across different layers, VGG11 (Fig.~\ref{fig:pea_vgg3}) showcases a promising latency-to-output data size ratio in layers 3, 6, 11, and 27. However, this dynamic slightly shifts with VGG19 (Fig.~\ref{fig:pea_vgg4}), where layers 5, 10, 19, and 43 demonstrate enhanced efficiency in output size, despite significant computational overhead before these potential cut points, as depicted in Figs.~\ref{fig:pea_vgg5} and~\ref{fig:pea_vgg6}.

\begin{figure}[ht]
        \centering
        \subfigure[VGG11]{
            \centering
            \includegraphics[width=4.52cm]{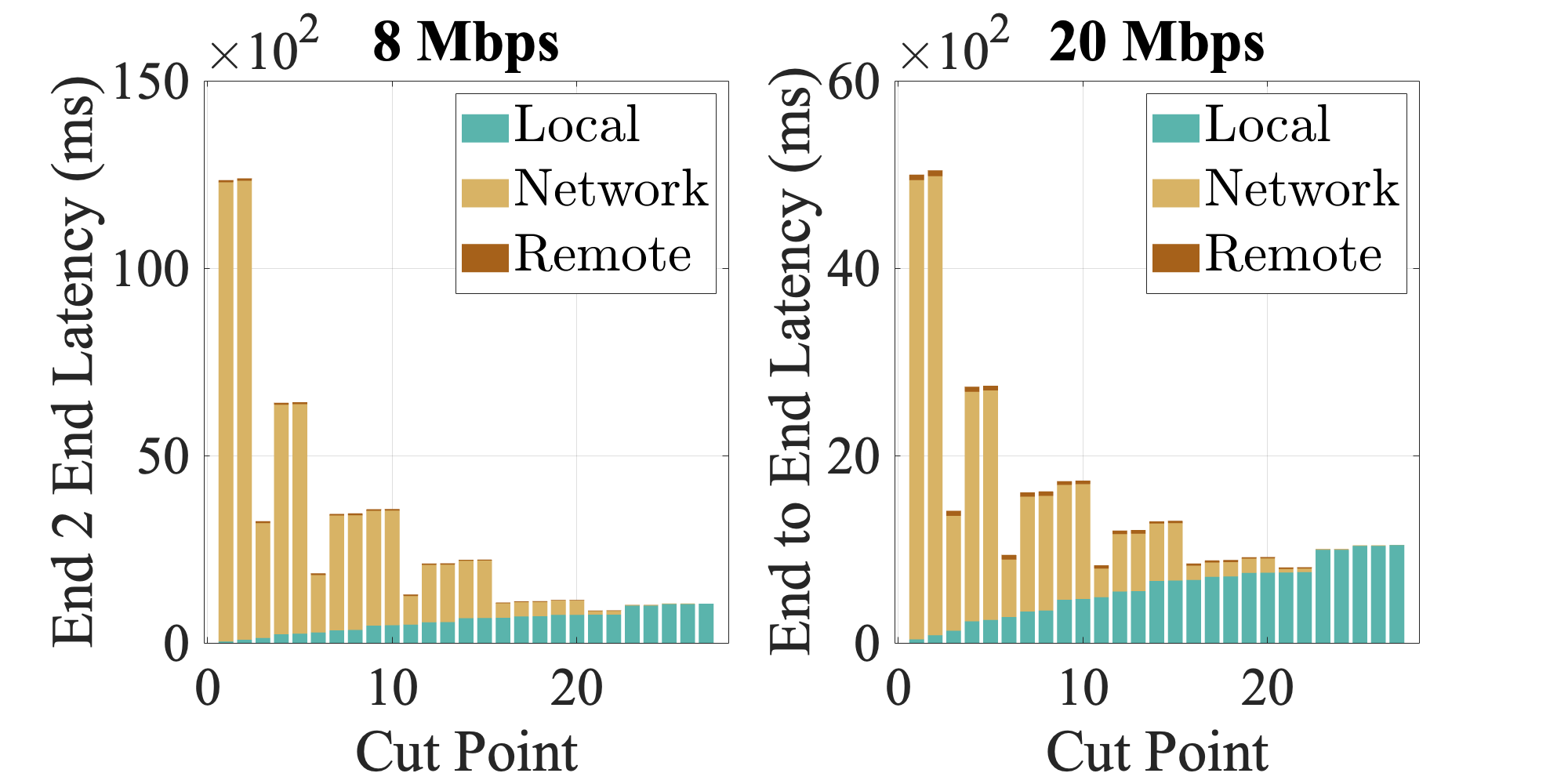}
        \label{fig:e2e_vgg11}
        }\hspace{-0.32in}
        \subfigure[VGG19]{
            \centering
            \includegraphics[width=4.52cm]{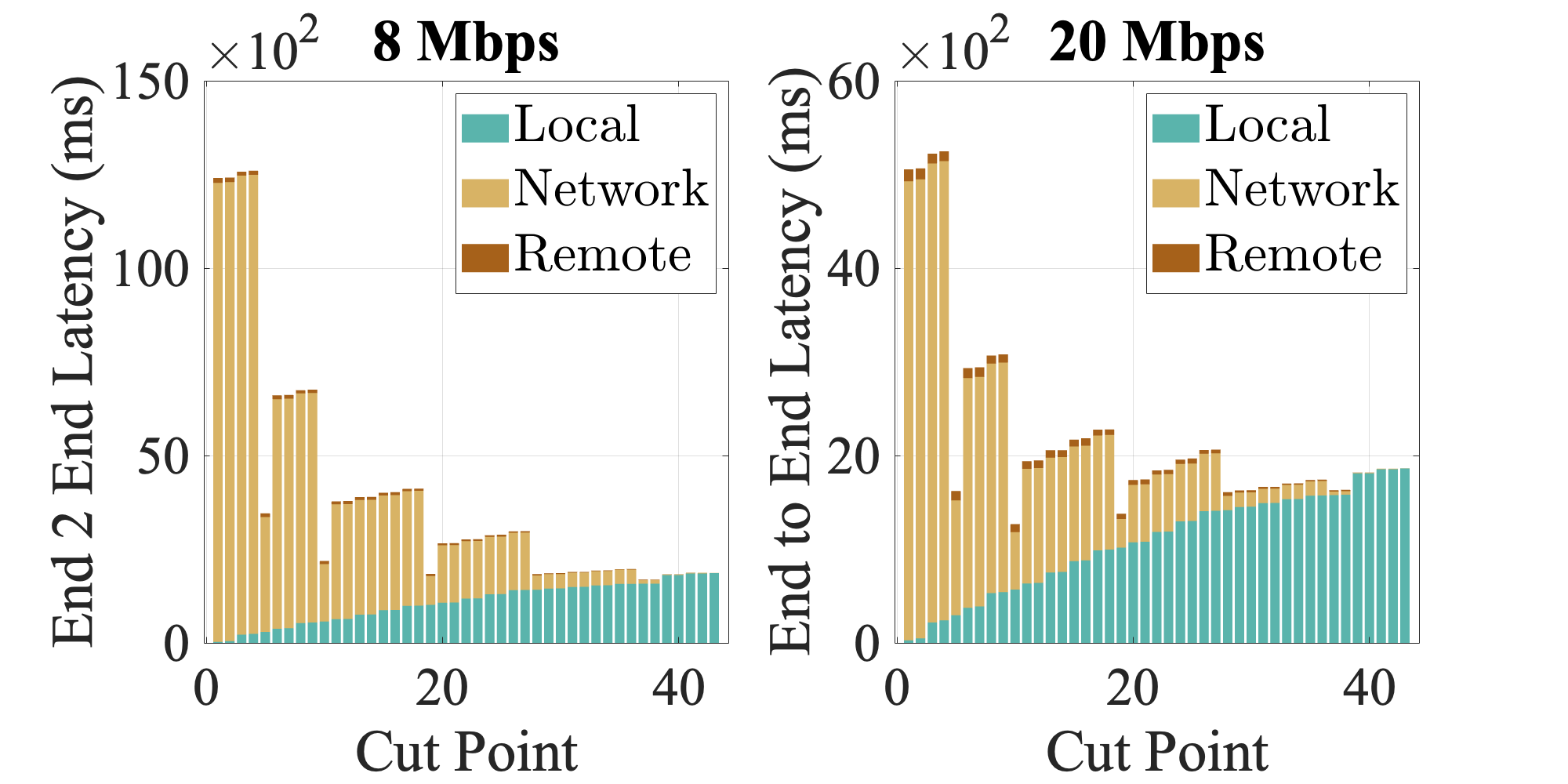}
            \label{fig:e2e__vgg19}
        }
        \vspace{-0.2in}
        \caption{\footnotesize{End-to-End latency comparison of different versions of VGG model} 
    }
    \label{fig:e2e}
     \vspace{-0.2in}
\end{figure}
        
\begin{figure}[ht]
        \centering
        \subfigure[\footnotesize{VGG11}]{
            \centering
            \includegraphics[width=4.52cm]{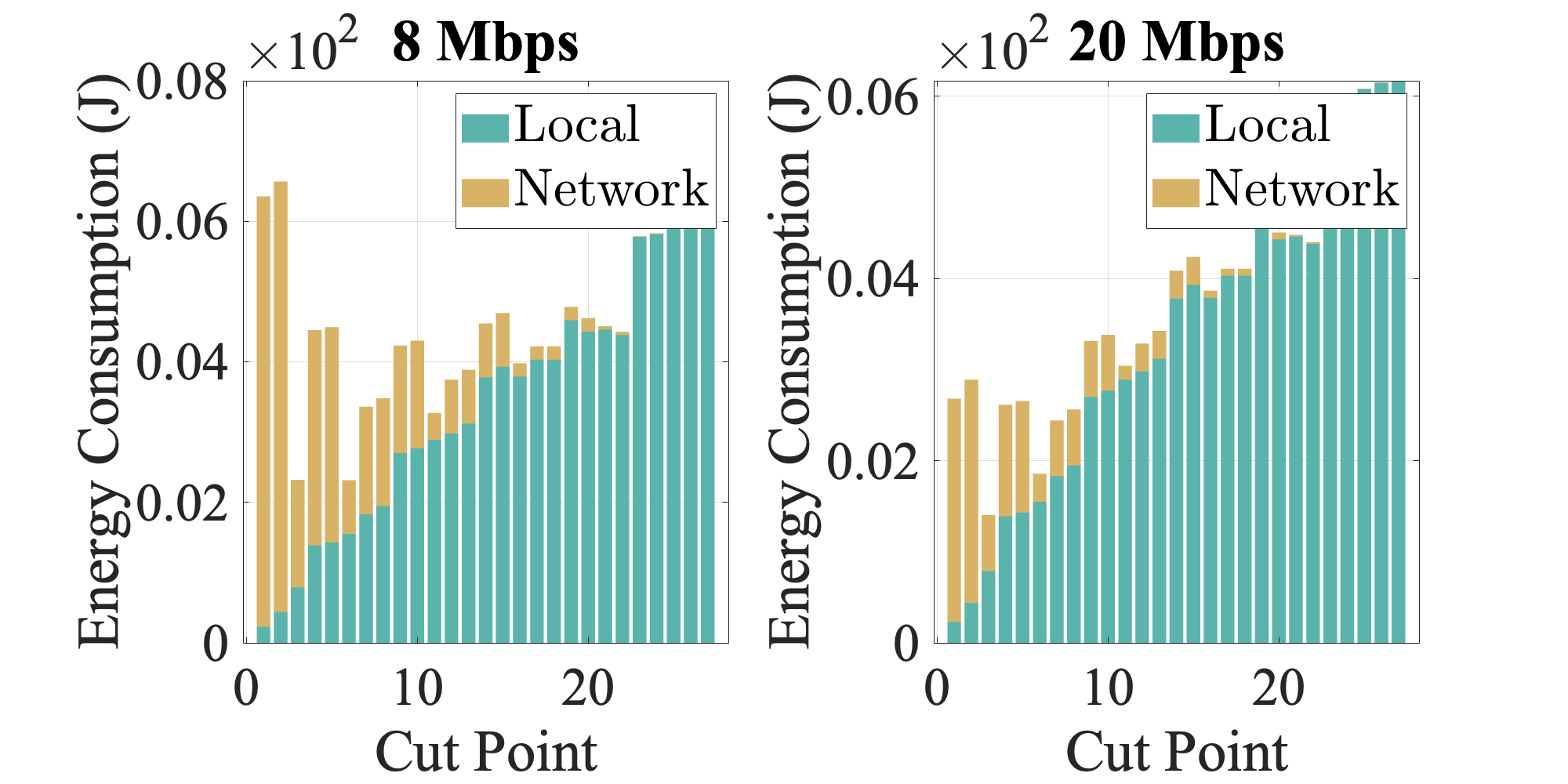}
            \label{fig:e2ee_vgg11}
        }\hspace{-0.32in}
        \subfigure[\footnotesize{VGG19}]{
            \centering
            \includegraphics[width=4.52cm]{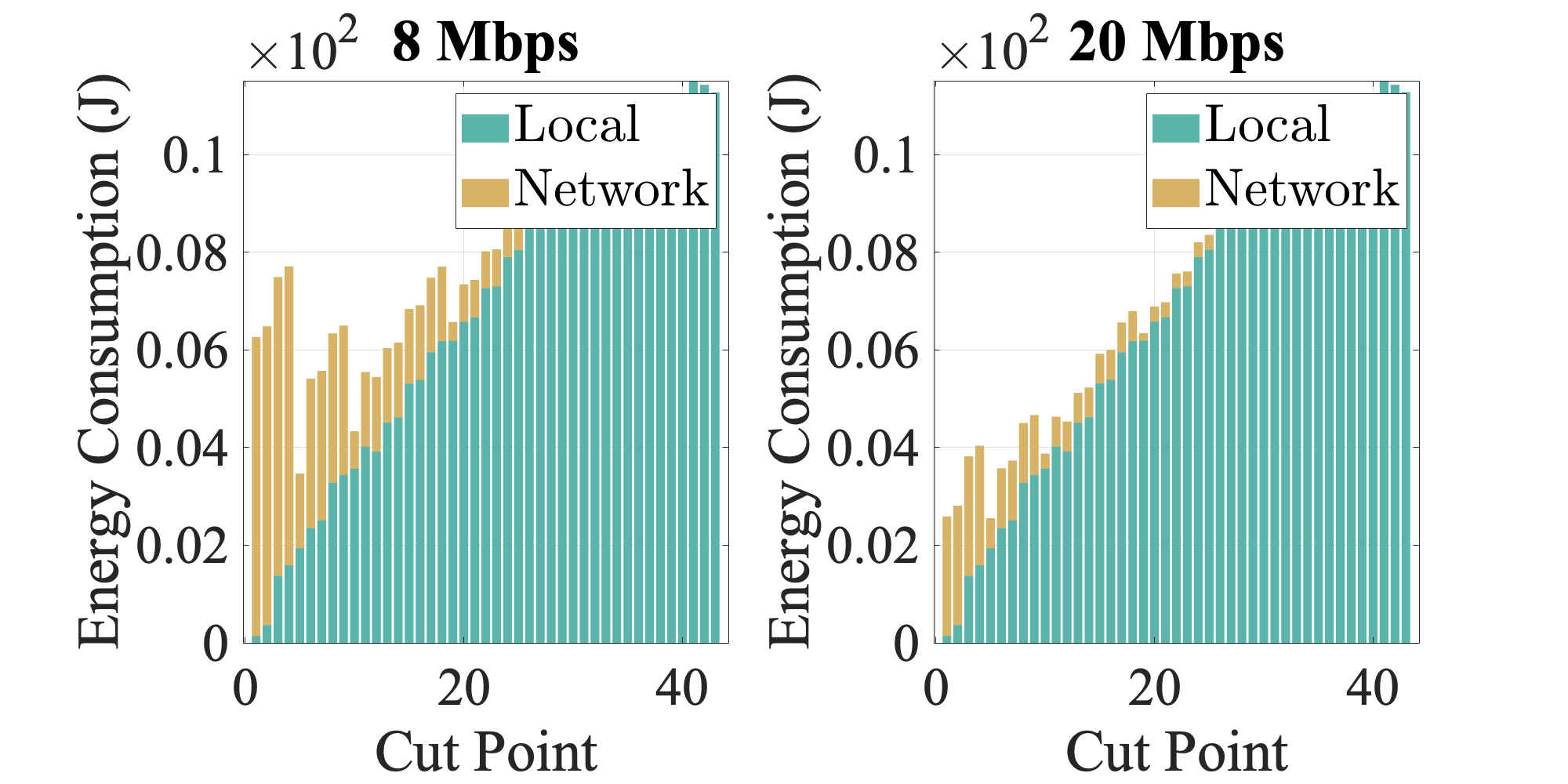}
            \label{fig:e2ee_vgg19}
        }
        \vspace{-0.2in}
        \caption{\footnotesize{Energy consumption comparison of different versions of VGG model} 
    }
    \label{fig:e2ee}
     \vspace{-0.1 in}
\end{figure}



\noindent \textbf{Layer-wise Characteristics.} Next, we measure end-to-end latency and energy consumption for all layers of VGG. End-to-end latency involves processing the initial part of the DNN, including the cut point layer, on the IoT device, transmitting the output of the cutpoint layer wirelessly, and subsequently resuming the inference task to the final layer on the server. For end-to-end latency, we exclude server energy consumption. Fig.~\ref{fig:e2e} 
summarizes the end-to-end latency and energy consumption of VGG11 and VGG19, 
over 8 Mbps and 20 Mbps transmission speeds, simulating LTE and WiFi connectivity. For VGG11, the optimal cutpoints in terms of end-to-end latency are layers 21 and 22. However, in terms of energy consumption, layers 6 and 3 are the best for 8 Mbps and 20 Mbps transmission speeds, respectively. Similarly, for VGG19, the optimal cut-points in terms of end-to-end latency are 19, 37, and 38 for 8 Mbps, and 10 and 19 for 20 Mbps. However, in terms of energy consumption, layer 5 is the best for both. Comparing the most efficient cut-points in terms of energy between versions, we can observe that the best cut-point of VGG19 uses almost twice the energy as VGG11.

{\em Considering these observations, we argue that `just-in-time' edge environments, handling latency-sensitive DNN inference, often deal with a trade-off involving end-to-end inference latency, device energy consumption, and model accuracy. While decisions, such as where to make the cut and which version to choose can significantly impact such trade-off, inherent dynamism in resource availability in `just-in-time' edge environments such decision making a challenging undertaking.} 
\section{System Model and Solution Approach}
\label{sec:systemmodel}
In this section, we describe `just-in-time' edge-AI system model, 
and the RL based 
strategy employed by the proposed {\em Infer-EDGE} framework.
\subsection{System Model}
\begin{figure}[t]
    \centering
    \includegraphics[width=\linewidth]{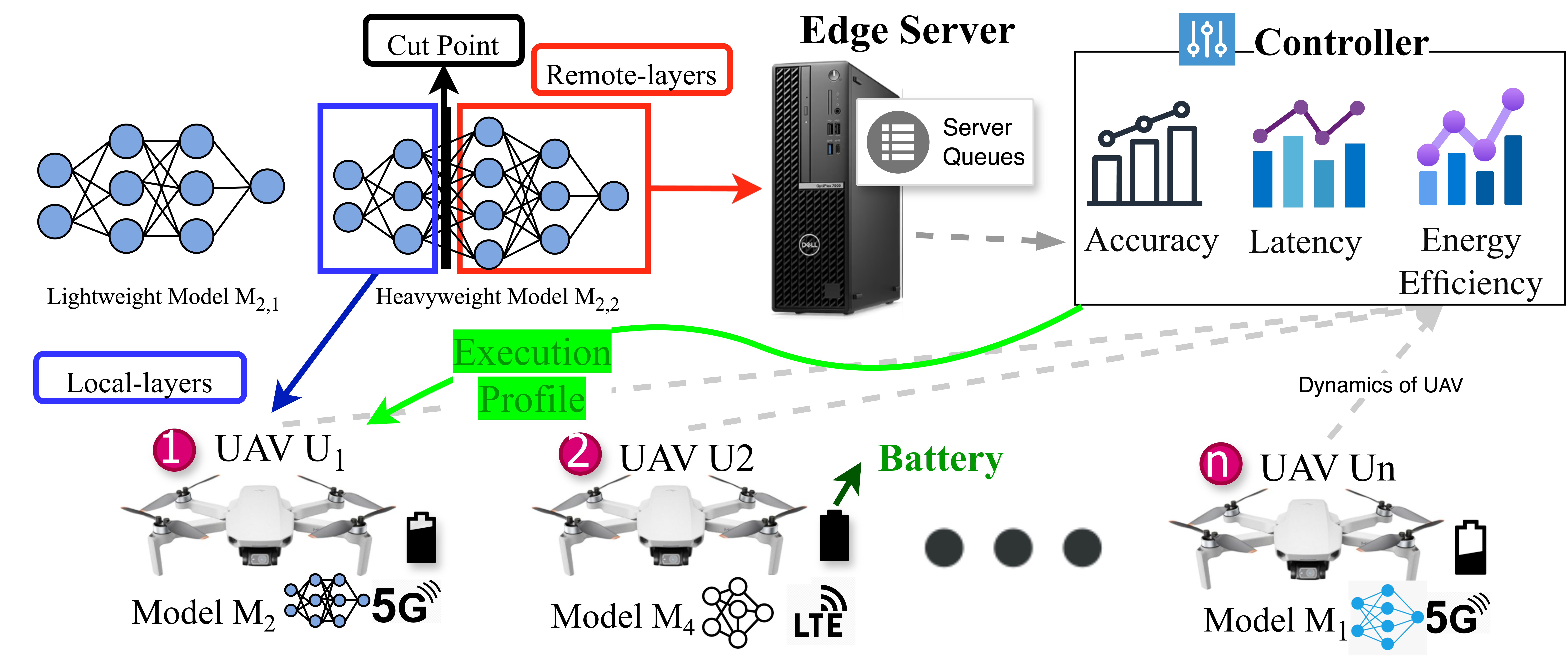}
    \vspace{-0.3in}
    \caption{\footnotesize{`Just-in-time' edge-AI system model for {\em Infer-EDGE} framework}}
    \vspace{-0.2in}
    \label{fig:systemmodel}
\end{figure}

%
As shown in Fig.~\ref{fig:systemmodel}, we assume an exemplary `just-in-time' edge environment where IoT devices (UAVs in this case), in collaboration with one limited capacity edge server, perform DNN model inference for real-time missions. We define a set of \textit{m} DNN models, denoted as $\mathcal{M} = \{M_1, M_2, ..., M_m\}$, each tailored for specific objectives and tasks towards the missions. 
We consider model \( M_i \) to have \( V_i \) different versions \( \{M_{i,1}, M_{i,2}, ..., M_{i,V_i}\} \), generated as a result of model optimization, with each employing either a compressed or extended architecture with diverse layers. These versions exhibit unique characteristics in accuracy and computational complexity.
The accuracy and number of layers of the \textit{i}-th model in its \textit{j}-th version are expressed as $M_{i,j}^{\text{acc}}$ and $M_{i,j}^{\text{layers}}$, respectively.
In this context, \( M_{i,j}^l \) denotes the `head' of the model up to and including layer \( l \) when referring to computations on the IoT device or local computation, and the `tail' of the model from layer \( l+1 \) onwards when referring to computations on the remote edge server.
%
For this work, we primarily consider video processing DNNs (e.g., object detection, object classification, object tracking) that are mostly used for mission critical use cases. However, the proposed {\em Infer-EDGE} framework solutions are universally adoptable for all classes of DNNs.
To be consistent with mission-critical use cases, we set stringent performance requirement for DNN model accuracy and latency.
Specifically, for each model $M_i$, it is mandated that the end-to-end inference latency must not exceed $\tau_i^{latency}$, and the accuracy must achieve at $\tau_i^{acc}$ for the underlying application to be successful.


\noindent{\bf Energy consumption model:} For the end-user IoT devices, we assume realistic scenarios and examples that are popular for mission critical use cases adopting `just-in-time' edge environment. Specifically, we model devices that perform other (i.e., mostly kinetic) activities on top of capturing video/image of a scene and partially performing computation. The objective is to create a realistic yet challenging device energy consumption scenario for the {\em Infer-EDGE} framework to address. 
To this end, we define a set of \textit{n} heterogeneous Unmanned Aerial Vehicles (UAVs) or drones (very common for mission critical use cases), denoted as $\mathcal{U} = \{U_1, U_2, ..., U_n\}$, each equipped with computational capabilities. 
The heterogeneity comes from the UAV model that defines UAV weights and architecture, battery level, and UAV kinetic activity profile, explained later.
Each UAV collaboratively execute a 
DNN inference task in collaboration with an edge server, facilitated through wireless connectivity between the two. 
Each UAV $U_k$ 
is defined by a quadruple \textit{(ID, build, battery level, trajectory
)},
where `ID' represents the UAV's unique identification, `build' specifies the UAV model, 
`battery level' reflects its current power status, and `trajectory' outlines the planned path.
In our model, there are three key sources of device energy consumption:
\vspace{-0.023in}
\begin{itemize}[leftmargin=*,itemsep=0pt]
    \item \textbf{Kinetic activity:} Each UAV has four distinctive motions in their flight path — forward movement, vertical ascent/descent, rotation, and hovering, with each generating varying energy consumption rates. For this work, we use the well accepted model in~\cite{stolaroff2018energy}.
    \item \textbf{Computation:} 
    The energy expenditure for locally processing the head of the model (e.g. \( M_{i,j}^l \)) at UAV \( U_k \) is: 
    \vspace{-0.05in}
    \begin{equation}
        E_{comp,i, j}^l(U_k) = P_{comp,i,j}(U_k) * T_{local,i,j}^l(U_k)
    \end{equation}
    \noindent where 
    $P_{comp,i,j}$ and $T_{local,i,j}^l$ represent 
    power consumption rate during the computation and the latency of executing $M_{i,j}^l$ at $U_k$, respectively.
    \item \textbf{Transmission:} This includes the energy consumed in wirelessly transmitting intermediate data generated after executing the cut point layer and the cut point information to the server (e.g., via WiFi or LTE).
    Such energy consumption is: 
    \vspace{-0.05in}
    \begin{equation}
        E_{trans,i,j}^l(U_k) = \beta_{k}(B) * D_{i,j}^l
        \vspace{-0.05in}
    \end{equation}
    where $\beta_k(\text{B})$ is transmission energy consumption rate with bandwidth $B$ and $D_{i,j}^l$ is the output data size at layer \textit{l}.

\end{itemize}
For simplicity, we ignore the energy consumption for video capture. Thus, the total UAV  energy consumption is:
\vspace{-0.05in}
\begin{equation}
    E_{i,j}^l(U_k) = E_{comp,i,j}^l(U_k) + E_{trans,i,j}^l(U_k)  
    \vspace{-0.05in}
\end{equation}
For this work, we assume that the edge server(s) is connected to a power source for the duration of the mission. Therefore, the energy consumption of server is excluded from the analysis.

\noindent{\bf End-to-end latency model:} As part of `just-in-time' edge environment, we assume edge server(s) (one as shown in Fig.~\ref{fig:systemmodel}) that are limited in capacity and often running multiple workloads involving multiple IoT devices as part of a variety of mission critical applications. Thus, for a variety of DNN workloads (from heterogeneous devices as part of different missions) involving large video datasets, in most cases, the server can only process the `tail' of the models, instead of the entirety of model layers.
The latency \(T_{i,j}^l\) for executing \(M_{i,j}\) collaboratively between \(U_k\) and the edge server $\mathcal{E}$, partitioned at cut point \(l\), includes the following components:
\begin{itemize}[leftmargin=*,itemsep=0pt]
    \item \textbf{Local processing time (\( T_{local,i,j}^l \)): } 
    Latency of processing the `head' of $M_{i,j}^l$ at \( U_k \). This is dictated by the limited processing cycle available at $U_k$.   
    \item \textbf{Transmission time (\(T_{trans,i,j}^l(U_k, \lambda)\)):} 
    Data transfer latency between $U_k$ and the edge server, dictated by the transmission rate, i.e., upload bandwidth \(\lambda\). Notably, such bandwidth can be very limited depending the underlying mission-critical use case and often proves to be a limiting factor against the case of fully offloading $M_{i,j}^l$ to the edge server, thus necessitating partial offloading~\cite{zhang2021effect}.
    \item \textbf{Server/remote processing time (\(T_{remote,i,j}^l\)):}
    This includes the computation time (\(T_{\text{comp},i,j}^l\)) for $M_{i,j}^l$'s tail on the server $\mathcal{E}$, plus the server queue time (\(T_{queue}\)).
\begin{equation}
    T_{remote,i,j}^l(\mathcal{E}) = T_{queue}(\mathcal{E}) + T_{comp,i,j}^l(\mathcal{E})
\end{equation}
The variability in \( T_{\text{queue}} \) at server \( \mathcal{E} \), influenced by concurrent tasks managed for other jobs by the server, is crucial for accurately modeling the operational dynamics of limited resource `just-in-time' edge servers that support multiple application workloads generated from the mission-critical use case.
\end{itemize}
Thus, the total end-to-end latency 
is:
\begin{equation}
    \scalebox{0.83}{$
    T_{i,j}^l(U_k, \lambda, \mathcal{E}) = T_{local,i,j}^l(U_k) + T_{trans,i,j}^l(U_k, \lambda) + T_{remote,i,j}^l(\mathcal{E})
    $} 
    \label{totallatency}
\end{equation}
%

\noindent{\bf Controller and system management:} On top of the UAVs and the edge server(s), one more system component that is responsible for DNN optimization related decision making is a centralized controller, as shown in Fig.~\ref{fig:systemmodel}.
Physically, the controller can be implemented within the edge server(s) or as a separate entity.
The controller plays a pivotal role by collecting critical resource availability data, specifically server workload (i.e., $T_{queue}(\mathcal{E})$) and available bandwidths (i.e., $\lambda$) from the server the UAVs respectively, to determine optimal execution profiles, including selecting model versions and cut-point layers for each UAV.
These execution profile decisions are then promptly communicated to the server and the UAVs, enabling them to initiate model execution based on the decision. As mentioned earlier, due to the dynamic nature of the operating environment of the use cases where `just-in-time' edge systems are adopted, the proposed iterative approach ensures dynamic adaptation to varying environmental conditions, optimizing system performance and resource utilization.

\subsection{A2C based RL Agent}
In {\em Infer-Edge}, the controller trains a DRL agent to manage the aforementioned dynamism of system availability in `just-in-time' edge environments. In particular, we formulate the overall DNN optimization problem as a Markov Decision Process (MDP) and deploy a timeslot-based decision-making approach grounded in the Advantage Actor-Critic (A2C) algorithm~\cite{konda1999actor,mnih2016asynchronous}.
The choice of A2C is driven by its efficiency and effectiveness in DRL. In A2C, an agent serves both as the actor and the critic, combining policy-based and gradient-based methods. The actor makes decisions, while the critic evaluates these decisions and provides feedback to refine strategies. This collaborative approach accelerates training and enhances learning with each experience. Moreover, A2C is a stable algorithm capable of handling large observation spaces, such as `just-in-time' edge environments with potentially multiple UAVs and corresponding DNN models and versions.
The A2C agent operates in an environment characterized by a finite set of states denoted as $\mathcal{S}$ and a finite set of actions denoted as $\mathcal{A}$, 
under a time-slot based system, 
with intervals of $\delta$ time units. The agent's observation space incorporates information on UAV's battery, the details of the task at hand, transmission bandwidth, and server resource availability. This knowledge encapsulates a comprehensive representation of the system's state at each time $t$.


\noindent \textbf{States:} \textit{S} denotes the state space of the environment. The state of our `just-in-time' edge system at time $t$, denoted by $s(t) \in S$, captures essential components such as current battery level of UAVs, their task availability, transmission bandwidth, the \textit{DNN} model, their activity profile, and server queue size during the upcoming timeslot. 
The activity profile of the UAV is defined as the distribution of forward flight, vertical movement, and rotational movement in the next $\delta$ seconds. The state of the environment at time $t$ is defined as:


\vspace{-0.15in}
\begin{flalign}
    \mathcal{S} =& \Big \{ 
    s(t) = [
    s_1(t), s_2(t), \ldots , s_n(t)] ~:~ \forall{k \in |\mathcal{U}|}, \nonumber \\
    &s_k(t) = (b_k(t), \alpha_k(t), P_k^t(t), m_k(t), F_k(t), V_k(t),
    R_k(t)), \nonumber \\
    & b_k(t) \in [1,10], \alpha_k(t) \in \{0,1\}
    \Big \} && \label{eq:state}
\end{flalign}
%
where $b_k(t)$ is the battery level of k\textsuperscript{th} UAV at time $t$, 
$\alpha_k(t)$ represents 
task availability, $P_k^t$ refers to available transmission power at $t$, $m_k$ defines the DNN model of the task, and finally $F_k(t)$, $V_k(t)$, and $R_k(t))$ denote the percentage of forward flight, vertical movement, and the rotational movements of the UAV for the task.

\noindent \textbf{Actions:}
The action space, denoted by $\mathcal{A}$, is a Multi-discrete action space comprises of the decision made for each UAV device.  At time $t$, the action $a(t)$ performed by the agent determines the execution profile, i.e., the DNN version and the cut point:
\vspace{-0.1in}
\begin{flalign}
    \mathcal{A} =& \Big \{  
    a (t) = [a_1(t), \ldots, a_n(t)] ~:~ \forall{k \in |\mathcal{U}|}, \nonumber \\
    &a_k(t) = (j, l), j \in V_{m_k(t)}\thinspace\text{and}\thinspace l \in L_{m_k(t),j}
    \Big \} && \label{eq:actions}
\end{flalign}
%
where $j$ is the version and $l$ is the cut point for model $M_{m_{k}(t)}$.

\noindent \textbf{Reward Function:}
The reward function $R(t)$ denotes the immediate reward acquired following the transition from state $s(t)$ to state $s(t + 1)$ by executing action $a(t)$. In {\em Infer-EDGE} framework's A2C model, we take a weighted average approach of the three system performance requirement metrics, viz., device energy expenditure, model accuracy, and end-to-end inference latency.
Such approach makes the solution flexible where different combination of the weights can be designated based on system objectives and the relative priorities of performance metrics. 
To this end, we define separate normalized performance scores for accuracy ($\mathcal{A}$), latency ($\mathcal{L}$), and energy consumption ($\mathcal{E}$). 


Given that UAV \( U_k \) is running DNN model \( M_i \), the average reward function for an action 
$a_k = (j, l)$ over all the devices
can be defined as:



\vspace{-0.15in}
\begin{flalign}
    \scalebox{0.83}{$
    R(t) = \frac{1}{|\mathcal{U}|}\sum_{k\in\mathcal{U}} 
        w_1\mathcal{A}(M_{i,j}) + w_2\mathcal{L}(M_{i,j}^l, U_k)
        + w_3\mathcal{E}(M_{i,j}^l, U_k) 
    $} && \label{eq:reward}
\end{flalign}
\vspace{-0.15in}

where $ \sum_{i=1}^3 w_i = 1$ and the normalized performance scores are defined as follows: 
\vspace{-0.1in}
\begin{equation}
    \mathcal{A}(M_{i,j}) = \frac{1}{1+e^{-p.(M_{i,j}^{acc}-q)}}    
\end{equation}
\vspace{-0.1in}
\begin{equation}
    \mathcal{L}(M_{i,j}^l, U_k) = 1 - \frac{T_{i,j}^l (U_k, \lambda, \mathcal{E})}{T_{local,i,j}^{L_i} (U_k)}
\end{equation}
\vspace{-0.1in}
\begin{equation}
    \mathcal{E}(M_{i,j}^l, U_k) = 
    1 - \frac{E_{i,j}^l (U_k)}{E_{i,j}^{L_i} (U_k)}
\end{equation}
\vspace{-0.3in}

\begin{algorithm}[t]
\small
\caption{A2C-based controller algorithm
}
\label{alg:a2c}
\parbox[t]{\dimexpr\linewidth-\algorithmicindent}{\textbf{Output:} Actor network ($\theta$)}

\parbox[t]{\dimexpr\linewidth-\algorithmicindent}{
\textbf{Initialize} Actor ($\theta$) and Critic ($\theta_v$) networks with random weights}

\begin{algorithmic}[1]
\For{episode=1 to $\mathcal{Z}$ (total episodes)}  
    \State $t \gets 0$,  $s \gets [0]*|s|$, transition buffer $\gets$ []
    \For{$u \in \mathcal{U}$}
        \State Reset battery to full capacity
        \State Update activity profile for $u \in \mathcal{U}$
    \EndFor
    \While{ Any device is on }
        \For{$u \in \mathcal{U}$}
            \State Update: battery level, DNN model and transmission power
            
        \EndFor
        \State \parbox[t]{\dimexpr\linewidth-\algorithmicindent-25pt}{Update state $s$ using Eq.~\eqref{eq:state}}
        \State \parbox[t]{\dimexpr\linewidth-\algorithmicindent-15pt}{Input state $s$ to the Actor with weights $\theta$ and get the action $a_t$ 
        according to policy $\pi(a|s; \theta)$}
        \State \parbox[t]{\dimexpr\linewidth-\algorithmicindent-25pt}{Receive reward $r_t$ (Eq.~\eqref{eq:reward}) and switch to state $s'$}
        \If{action $a_t$ is valid}
            \State \parbox[t]{\dimexpr\linewidth-\algorithmicindent-25pt}{ Execute profiles based on (version, cut points) from $a$}
        \EndIf
        \State Append $(s_t, a_t, r_t)$ to transition buffer \textcolor{red}{$\beta$}
        \State $t \gets t + \delta$
        \State $s \gets s'$
    \EndWhile
    \State \parbox[t]{\dimexpr\linewidth-\algorithmicindent-25pt}
    {Compute discounted rewards \(\hat{R}_t\) and advantages using \(A(s,a) = Q(s,a) - V(s)\)}
    \State Update $\theta$ by policy gradient and $\theta_v$ by mean square error
    

\EndFor
\end{algorithmic}
\end{algorithm}

\subsection{A2C Algorithm and Architecture}
The 
(A2C)~\cite{mnih2016asynchronous} model is a hybrid method comprises of two neural networks: i) The Actor network which is a Policy Gradient algorithm that learns a policy $\pi$ 
 deciding on what action to take, and ii) the Critic network which is a Q-learning algorithm offering feedback for policy enhancement.  
 For {\em Infer-EDGE}, we design an online learning algorithm (as seen in Algo.~\ref{alg:a2c}) which runs on the centralized controller working as the system manager. 
 During initialization phase, the agent creates actor and critic networks with randomly assigned weights. The agent then continuously interacts with the environment and makes execution profile decisions in each time slot ($\delta$). At the end of each episode, both actor and critic networks' weights undergo weight updates with a batch of experienced transitions. Our Critic network features two fully connected layers with feature sizes of 512 and 256, respectively. To adapt the Multi-Discrete action structure, which determines the version and cut-point for each device, the actor network incorporates an additional shared layer for related action values. Specifically, every two values that correspond to each UAV device share an extra layer with a feature size of 128.

\subsection{Framework Design}
\begin{figure}[t]
    \centering
    \includegraphics[width=0.8\linewidth]{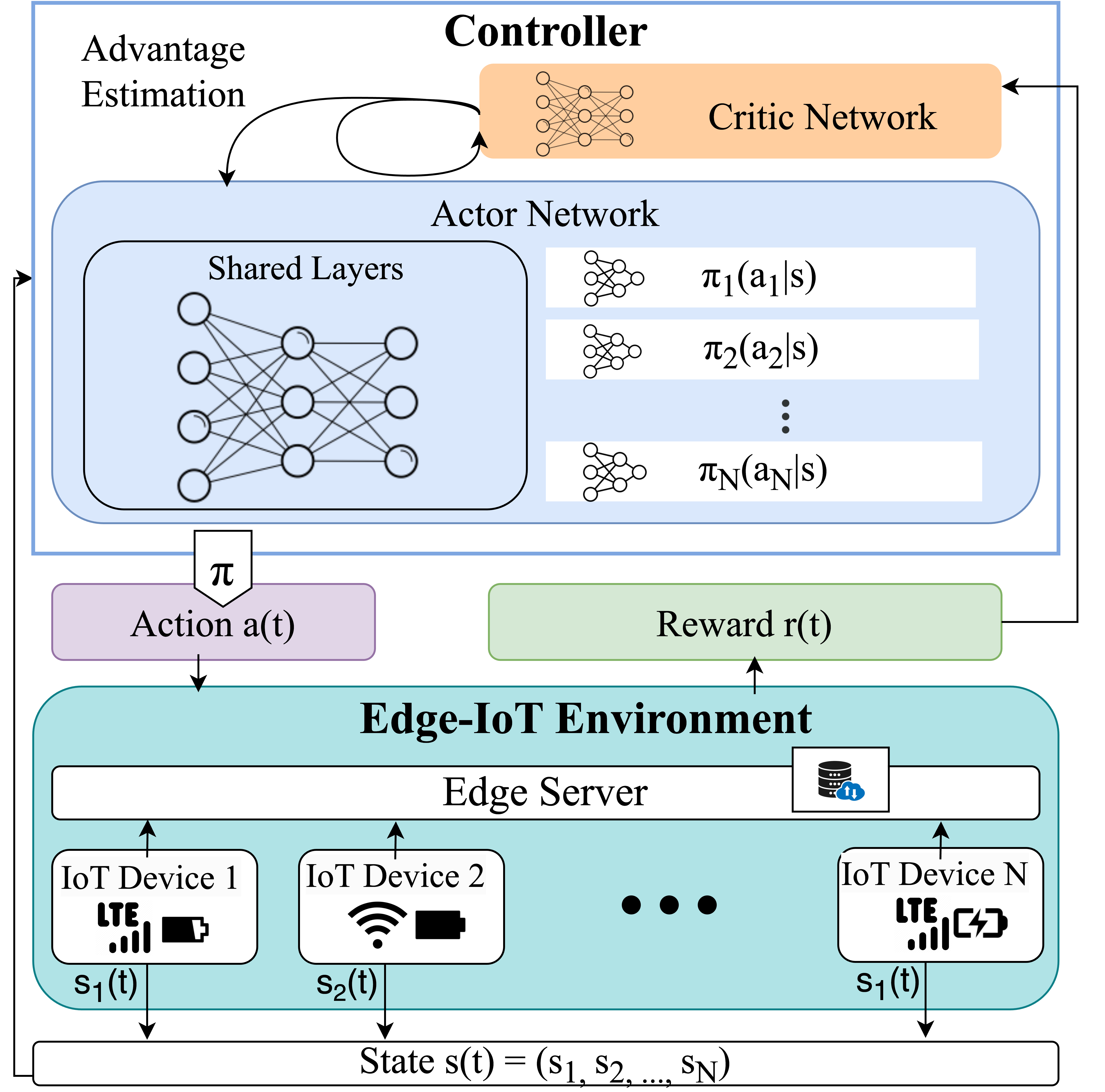}
    \vspace{-0.1in}
    \caption{{\em Infer-EDGE} framework with centralized controller implementing the proposed A2C algorithm}
    \label{fig:enter-label}
    \vspace{-0.2in}
\end{figure}

Fig.~\ref{fig:enter-label} illuistrates {\em Infer-Edge} framework design with a centralized controller, serving as the system manager for decision-making processes. The UAV devices play a crucial role by transmitting essential information such as task details, battery levels, and available transmission speeds to the controller. This aggregated data forms the system's state, which is then processed by an actor network within the controller. The actor network utilizes this information to generate actions, taking into account various factors like system performance and resource availability.
Once actions are generated, they are relayed back to the respective UAV devices. Simultaneously, the system records rewards based on performance metrics such as accuracy, latency, and energy consumption within the `just-in-time' edge environment. Following this, a critic network estimates the advantage values and trains both the actor and critic networks based on the actions taken and the resulting rewards.
Continuing through this iterative learning process, the system refines and adapts until it reaches convergence, ensuring optimal performance and responsiveness to environmental variables. Each episode concludes as all UAV device batteries are depleted.

\section{Evaluation}\label{evaluaton}


Next, we evaluate the performance of our proposed {\em Infer-EDGE} framework through hardware testbed experimental evaluation. 
%
\subsection{Testbed Setup and Experiment Design}
For our `just-in-time' edge testbed, we utilize three NVIDIA Jetson TX2 devices, viz., `Aruna Ali,' `Valentina Tereshkova,' and `Malala Yousafzai' — as computational units of the IoT devices/UAVs. Additionally, a Dell PowerEdge desktop with 16 cores 3.2 GHz CPU, viz., `Grace Hopper', serves as the edge server. 
The network connectivity between the TX2 devices and the edge server is established through an Ettus USRP B210 acting as the access point, and can operate on both WiFi and LTE bands.  
Due to lack of UAV hardware availability, we simulate UAV kinetic activity based on an average size drone \textit{UAV Systems Aurelia
X4 Standard} and compute energy consumption of each movements based on the model proposed in~\cite{stolaroff2018energy}. In order to add device heterogeneity, we define three \textit{activity profiles} with varying coverage levels (Tab. \ref{tab:movement_distribution}), where a greater forward flight rate is generally equivalent to greater coverage. 

\begin{table}[t]
\small
\caption{Distribution of UAV kinetic activity}
 \vspace{-0.1in}
    \centering
    \begin{tabular}{l p{1.1cm} p{1.1cm} p{1.2cm}}
    \hline
    \textbf{Activity Level} & \textbf{Forward} & \textbf{Vertical} & \textbf{Rotational} \\
    \hline
    High & 80\% & 10\% & 10\% \\
    Moderate & 50\% & 25\% & 25\% \\
    Low & 20\% & 40\% & 40\% \\
    \hline
    \end{tabular}
    \vspace{-0.2in}
    \label{tab:movement_distribution}
\end{table}
%

For the experiments, we mostly focus on object classification tasks as exemplar video processing applications. The UAV devices, thus execute three popular classification DNNs, viz., VGG, ResNet, and DenseNet.
As for different versions, as an outcome of DNN model architecture optimization, we assume that each DNN has two variants: a lightweight, less accurate model (e.g., VGG11, ResNet18, and DenseNet121), and a heavyweight, more accurate model (e.g., VGG19, ResNet50, and DenseNet161), as detailed in Table~\ref{tab:benchmarking}. Furthermore, drawing from insights presented in Section~\ref{sec:motivation}, we identify four potential cut points for each such DNN version (Table~\ref{tab:cutpoints1}), 
to enable collaborative DNN inference. 

\begin{table}[t]
\small
\caption{Candidate cut points for the each model}
\vspace{-0.1in}
    \centering
    \begin{tabular}{lcl}
        \hline
        \textbf{Model} & \textbf{Version} & \textbf{Candidate Cut Points} \\
        \hline
        \multirow{2}{*}{VGG} & 11 & 3, 6, 11, 27 \\
        & 19 & 5, 10, 19, 43 \\
        \hline
        \multirow{2}{*}{ResNet} & 18 & 4, 15, 20, 49 \\
        & 50 & 4, 13, 20, 115 \\
        \hline
        \multirow{2}{*}{DenseNet} & 121 & 4, 6, 8, 14 \\
        & 161 & 4, 6, 8, 14 \\
        \hline
    \end{tabular}
    \vspace{-0.2in}
    \label{tab:cutpoints1}
\end{table}

Apart from the object classification jobs arriving from the UAVs, we simulate the edge server to support other mission related jobs with exponential arrival rate which impact the size of the queue to follow a Poisson point process. 
For the classification jobs from UAVs, we employ a time slot duration of $\delta = 30s$, tailored to suit the specific demands of the reconnaissance missions employed by `just-in-time' edge environments. 

%

%
\vspace{-0.1in}
\subsection{Learning Stability}
\begin{figure}[t]
    \centering
    \subfigure[]{
        \includegraphics[width=3.8cm]{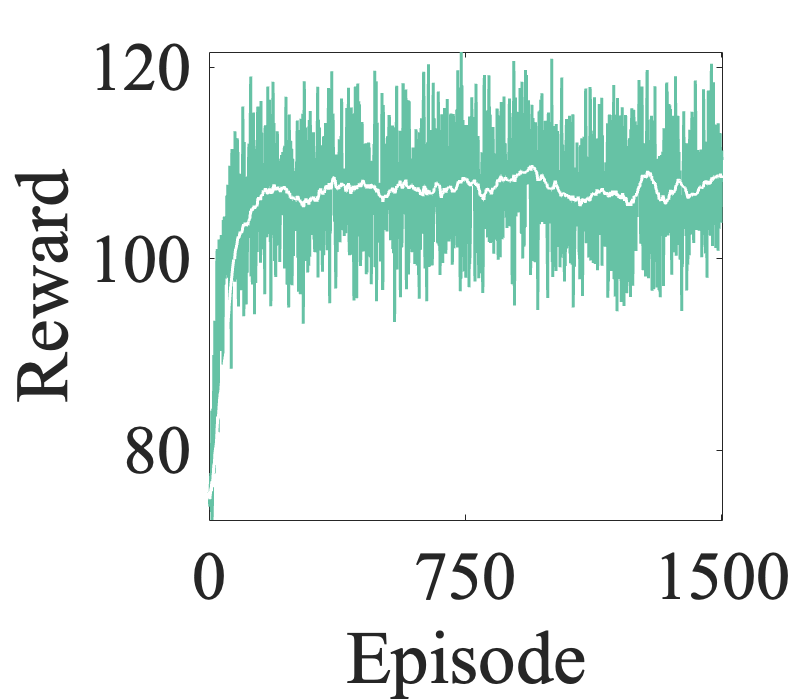}
        \label{fig:3convergence}
    }\hspace{-0.1in}
    \subfigure[]{
        \includegraphics[width=3.8cm]{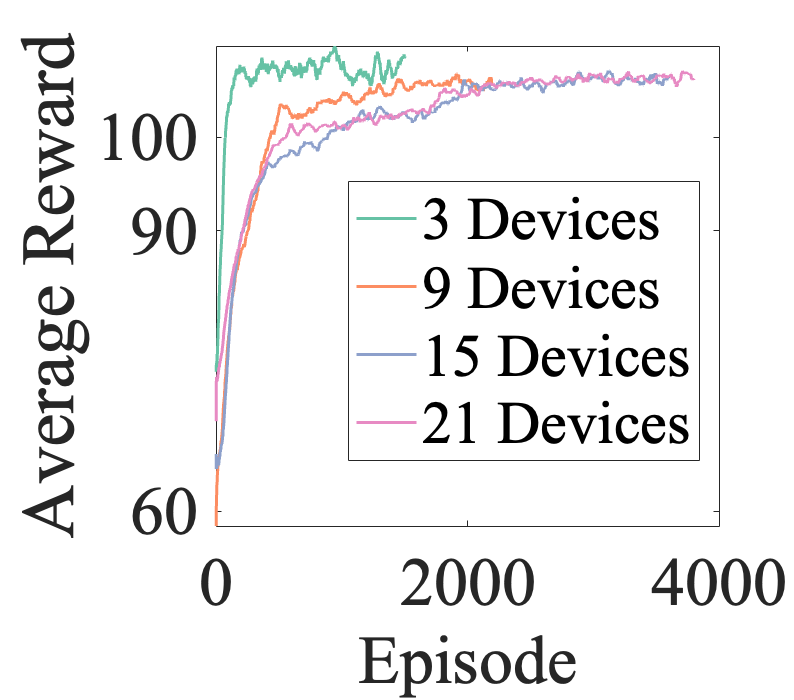}
        \label{fig:learningcurve}
    }
    \vspace{-0.1in}
    \caption{Average reward per episode with A2C algorithm}
    \label{fig:convergence}
    \vspace{-0.2in}
\end{figure}

Our RL model is trained on three UAVs over 5000 episodes, utilizing a learning rate of $5e^{-5}$. Throughout the learning process, the selection of activity profile, DNN model, and transmission speed are randomized, ensuring exposure to a diverse range of scenarios for comprehensive learning.
Fig,~\ref{fig:3convergence}) demonstrates the learning dynamics with changes in the number of UAVs as the convergence criterion. 
The figure shows that despite a slightly slower learning rates with increased UAVs, our system achieves convergence successfully. 
The results confirm the efficacy and stability of the proposed learning process, even with large number of devices. Despite a slight increase in learning episodes, performance remains robust, emphasizing the system's capability to handle larger `just-in-time' edge environments.

\subsection{RL Model Ablation study}
Next, we assess {\em Infer-EDGE}'s performance under different optimization strategies, specifically targeting latency only (viz., LO), accuracy only (viz., AO), and energy consumption only (EO). This involves adjusting the weights in Equation~\ref{eq:reward} to optimize each metric individually. This allows us to assess the individual impact of each factor on the proposed RL model's behavior. 

\subsubsection{Tri-factor performance:} 
\begin{figure}[t]
    \centering
    \subfigure[]{
        \centering
        \includegraphics[width=2.8cm]{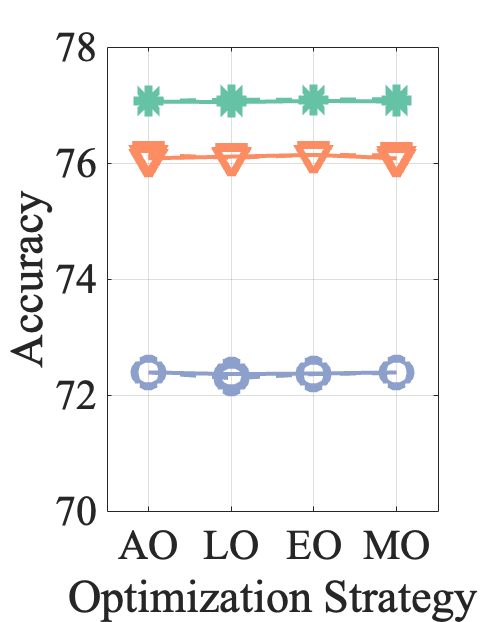}
        \label{fig:baseline_acc}
    }\hspace{-0.15in}
    \subfigure[]{
        \centering
        \includegraphics[width=2.8cm]{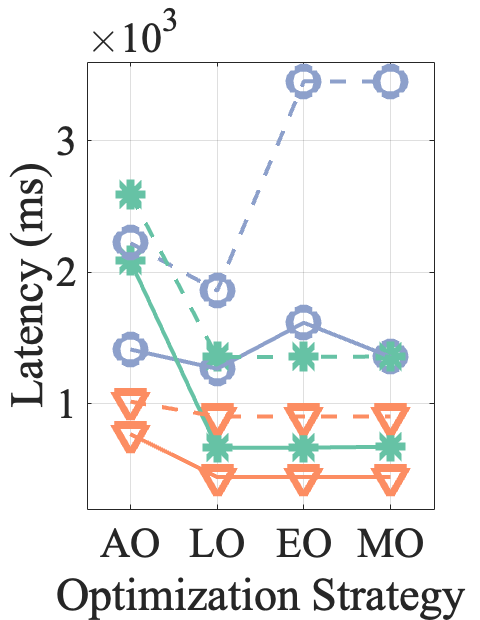}
        \label{fig:baseline_latency}
    }\hspace{-0.15in}
    \subfigure[]{
        \centering
        \includegraphics[width=2.8cm]{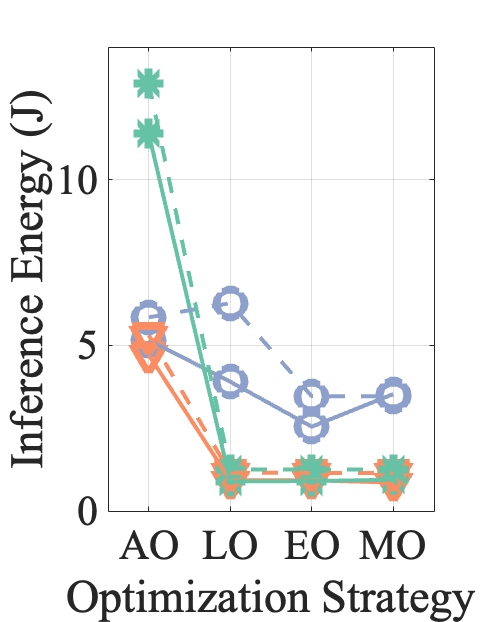}
        \label{fig:baseline_energy}
    }\hspace{-0.1in}
    \\
    \vspace{-0.15in}
    \subfigure{
        \centering
        \includegraphics[width=7cm]{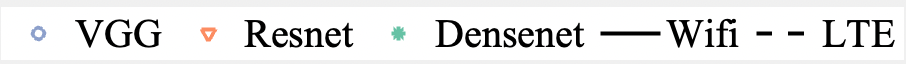}
    }
    \vspace{-0.1in}
    \caption{Performance comparison between RL strategies}
    \vspace{-0.1in}
    \label{fig:solobaseline}
\end{figure}

Here, we demonstrate the performance comparison of {\em Infer-EDGE}'s multi-objective (viz., MO) approach against baseline univariate AO, LO, and EO models for different data rate conditions, i.e., for WiFi (20 Mbps upload) and LTE (8 Mbps upload).
As shown in Figure \ref{fig:baseline_acc}, there is not much difference between univariate and multivariate models in terms of accuracy. As it turns out, the optimal cut points of the heavy-weight models are more efficient in terms of energy and latency, resulting in very close accuracy results. For latency optimization, proposed MO strategy performs significantly better than AO and Eo models. 
Similarly, for energy optimization, MO performs as good as EO, across different network condisions.

\subsubsection{Cut point layer selection:} 
Cut point layer selection comparison between univariate and multivariate RL models strategies for different upload data rate conditions are demonstrated in Tab.~\ref{tab:cutpoints}. We observe that in ResNet and DenseNet models, the first cut point is consistently favored regardless of the data rate. In contrast, for VGG models, particularly under lower data rates such as LTE, {\em Infer-EDGE}'s MO strategy partitons the DNNs until later layers to effectively reduce the data size, thereby mitigating transmission time. AO strategy results are omitted as its selection of cut points is entirely arbitrary and lacks meaningful interpretation.


\begin{table}[t]
\small
  \centering
  \caption{DNN cut point selection comparison}
  \vspace{-0.1in}
  \label{tab:cutpoints}
  \begin{tabular}{llccc}
    \hline
    Transmission Power & DNN & LO & EO & MO\\
    \hline
    \multirow{3}{*}{LTE} & VGG & 19 & 5 & 5 \\
                          & ResNet & 4 & 4 & 4 \\
                          & DenseNet & 4 & 4 & 4 \\
    \hline
    \multirow{3}{*}{WiFi} & VGG & 10 & 5 & 10 \\
                           & ResNet & 4 & 4 & 4 \\
                           & DenseNet & 4 & 4 & 4 \\
    \hline
  \end{tabular}
  \vspace{-0.2in}
\end{table}


\subsubsection{Latency/Energy savings}
Latency improvement and energy-saving performance comparisons across various strategies are demonstrated in Tab.~\ref{tab:lteimprovement}, for different network conditions, i.e., LTE and WiFi based upload speeds. 
The proposed MO strategy achieves energy saving similar to the EO strategy, while closely tracking the LO strategy in terms of latency improvement.
The above results, overall, shed light on the trade-offs and performance implications associated with optimizing the system for each individual performance metric. Such findings help in making informed decisions regarding parameter tuning in real-world `just-in-time' edge deployments.

\begin{table}[t]
\scriptsize
    \centering
    \caption{Latency and energy performance improvements}
    \vspace{-0.1in}
    \begin{tabular}{lcccccccc}
        \hline
        \multirow{2}{*}{Model} &
        \multicolumn{2}{c}{AO} &
        \multicolumn{2}{c}{LO} &
        \multicolumn{2}{c}{EO} &
        \multicolumn{2}{c}{MO}\\
         &LTE & WiFi &LTE & WiFi &LTE & WiFi &LTE & WiFi\\
        \hline
        Latency & 46\% & 60\% & 62\% & 78\% & 47\% & 75\% & 47\% & 77\%\\
        Energy & 65\%& 69\% & 87\%& 91\% & 91\%& 91\% & 91\%& 92\%\\
        \hline
    \end{tabular}
    \label{tab:lteimprovement}
    \vspace{-0.2in}
\end{table}

\subsection{Reward Sensitivity Analysis}
Next, we explore the sensitivity of the reward function for each 
performance metric 
across different DNN models. 


\subsubsection{Sensitivity of accuracy weight:}
\begin{figure}[t]
    \centering
    \subfigure[]{%
        \includegraphics[width=1.75cm]{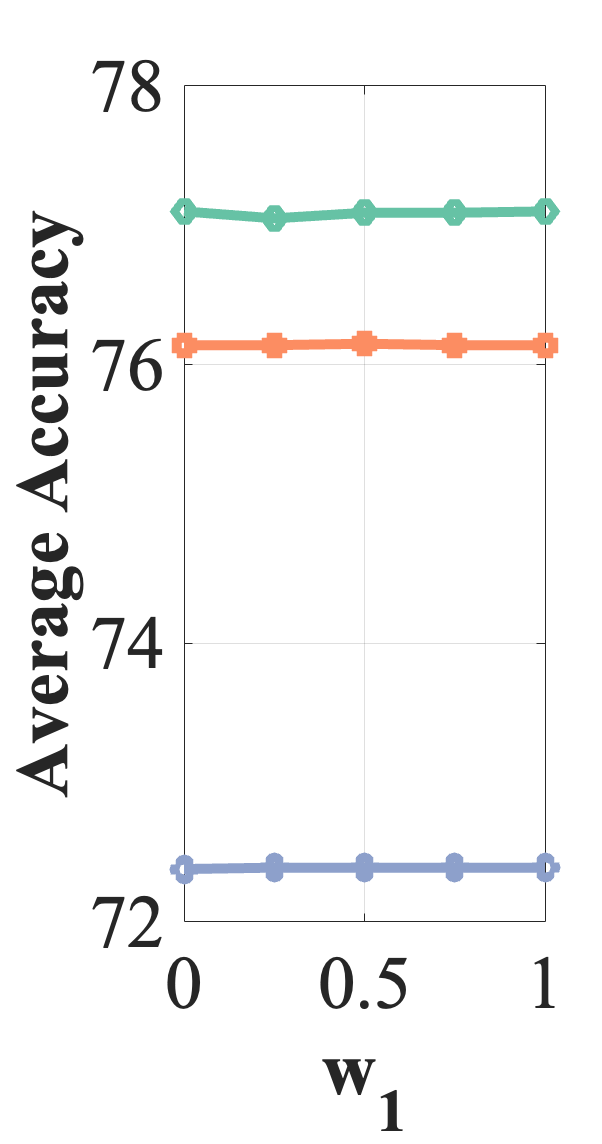}
        \label{fig:accuracy_accuracy}
    }\hspace{-0.1in}
    \subfigure[]{%
        \includegraphics[width=1.75cm]{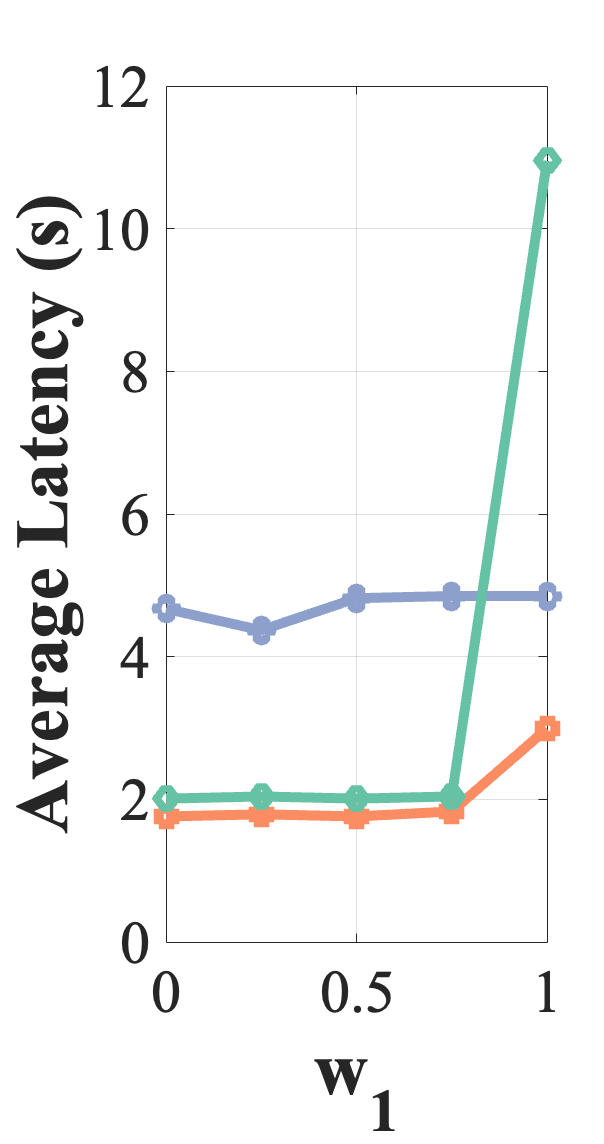}
        \label{fig:latency_accuracy}
    }\hspace{-0.1in}
    \subfigure[]{%
        \includegraphics[width=1.75cm]{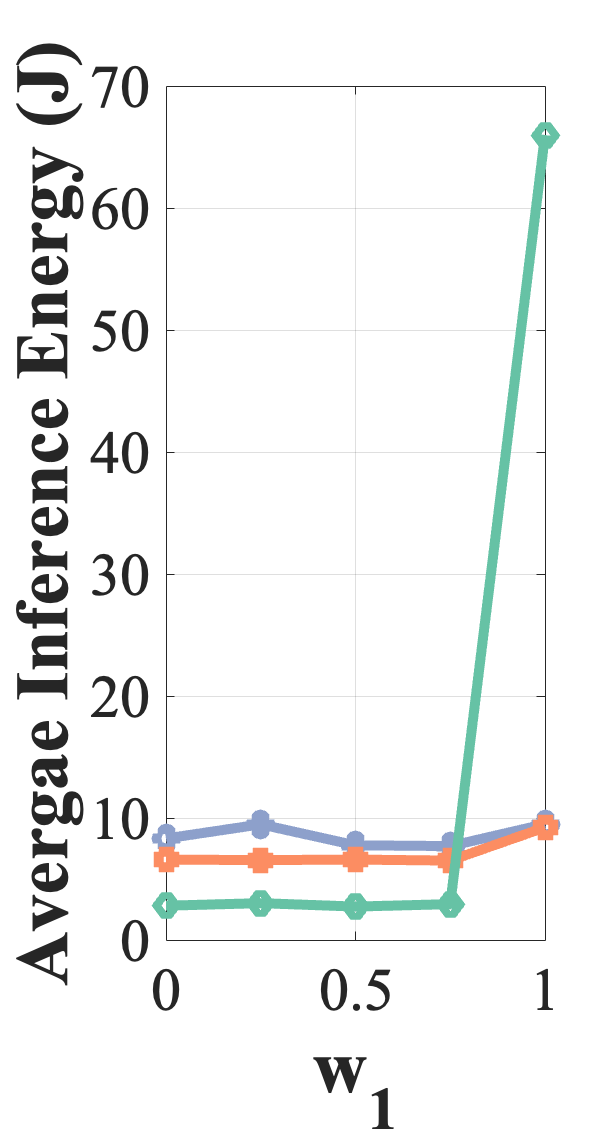}
        \label{fig:inference_accuracy}
    }\hspace{-0.1in}
    \subfigure[]{%
        \includegraphics[width=1.75cm]{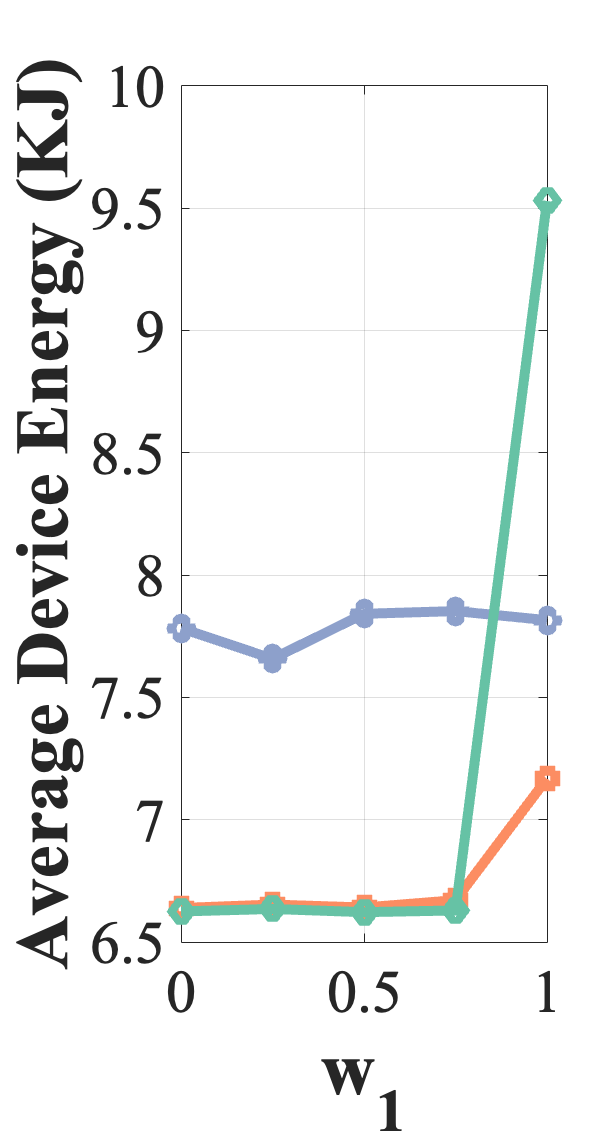}
        \label{fig:energy_accuracy}
    }\hspace{-0.1in}
    \subfigure[]{%
        \includegraphics[width=1.8cm]{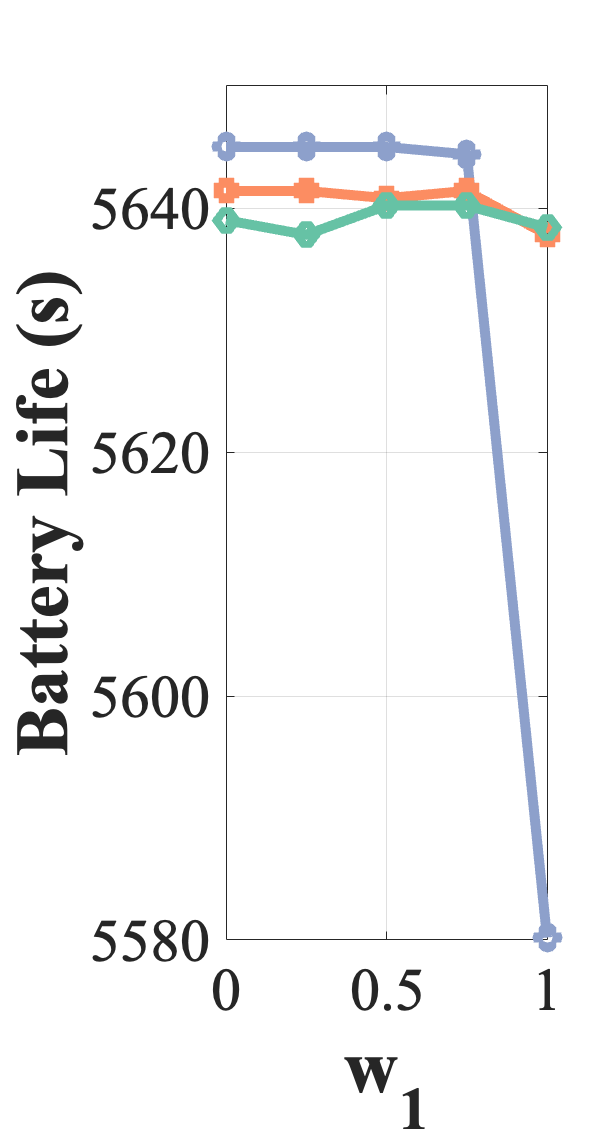}
        \label{fig:batterylife_accuracy}
    }\\
    \vspace{-0.15in}
    \subfigure{%
        \includegraphics[width=3cm]{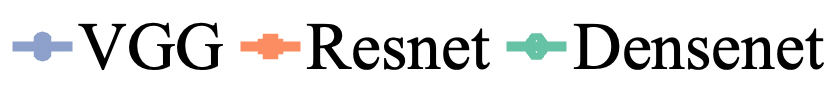}
    }\\
    \vspace{-0.1in}
    \caption{System performance over varying accuracy weight}
    \vspace{-0.2in}
    \label{fig:accuracyreward}
\end{figure}

Fig.~\ref{fig:accuracyreward} showcases the system performance for varying weight of accuracy reward in Eqn.~\eqref{eq:reward}. 
A notable observation is that (as seen in Fig.~\ref{fig:accuracy_accuracy}) the higher accuracy versions demonstrate better latency and energy efficiency. This is evident by the sustained high accuracy even when the accuracy reward weight is set to $0$. There is minimal improvement in accuracy with the increase in accuracy weight in the reward function. Moreover, as we increase the weight, there is a noticeable decline in latency and energy performance. This trend can be attributed to the selection of cut points.

\subsubsection{Sensitivity of latency weight:}
\begin{figure}[t]
    \centering
    \subfigure[]{%
        \includegraphics[width=1.75cm]{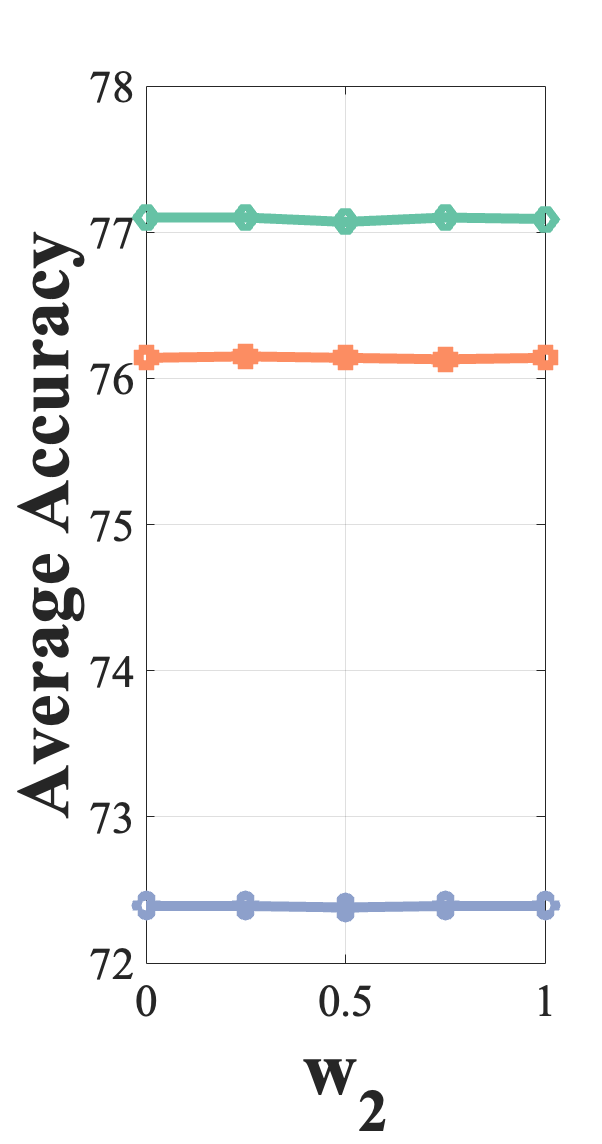}
        \label{fig:accuracy_latency}
    }\hspace{-0.1in}
    \subfigure[]{%
        \includegraphics[width=1.75cm]{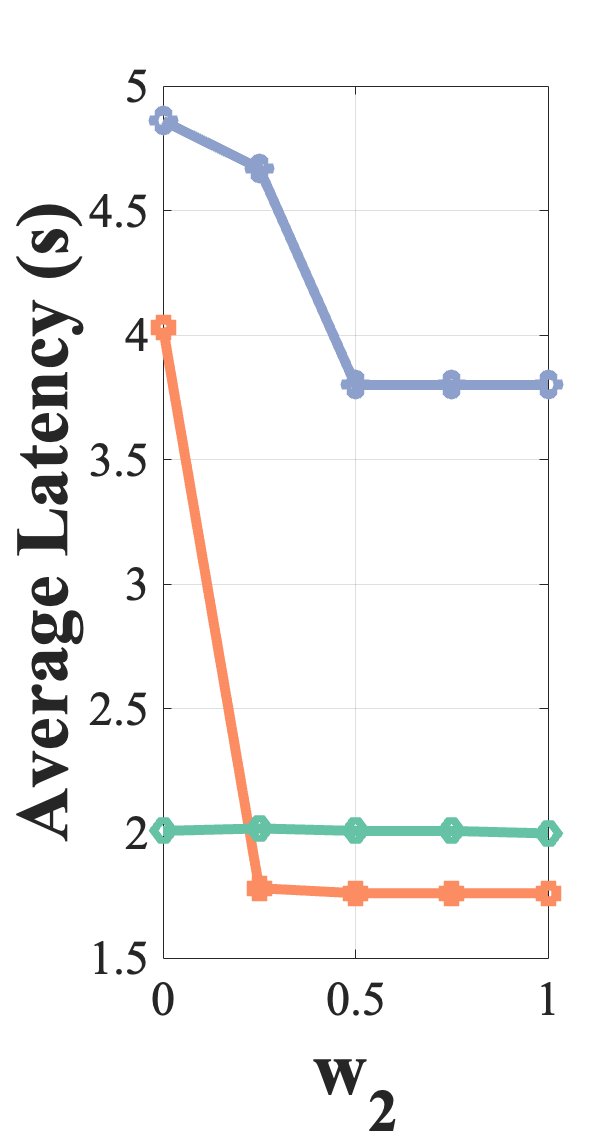}
        \label{fig:latency_latency}
    }\hspace{-0.1in}
    \subfigure[]{%
        \includegraphics[width=1.75cm]{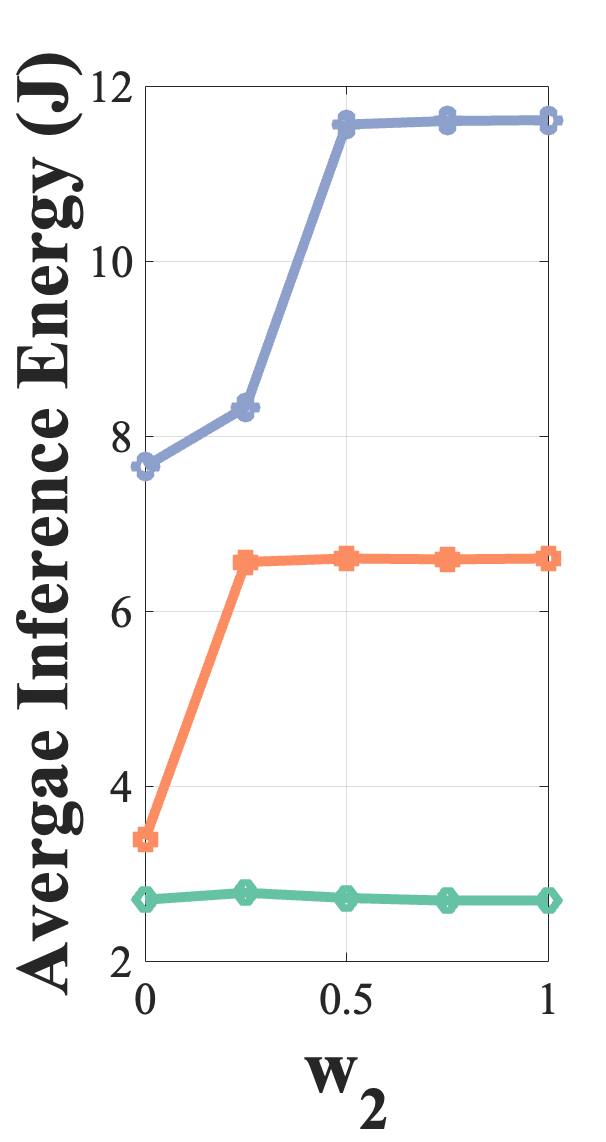}
        \label{fig:inference_latency}
    }\hspace{-0.1in}
    \subfigure[]{%
        \includegraphics[width=1.75cm]{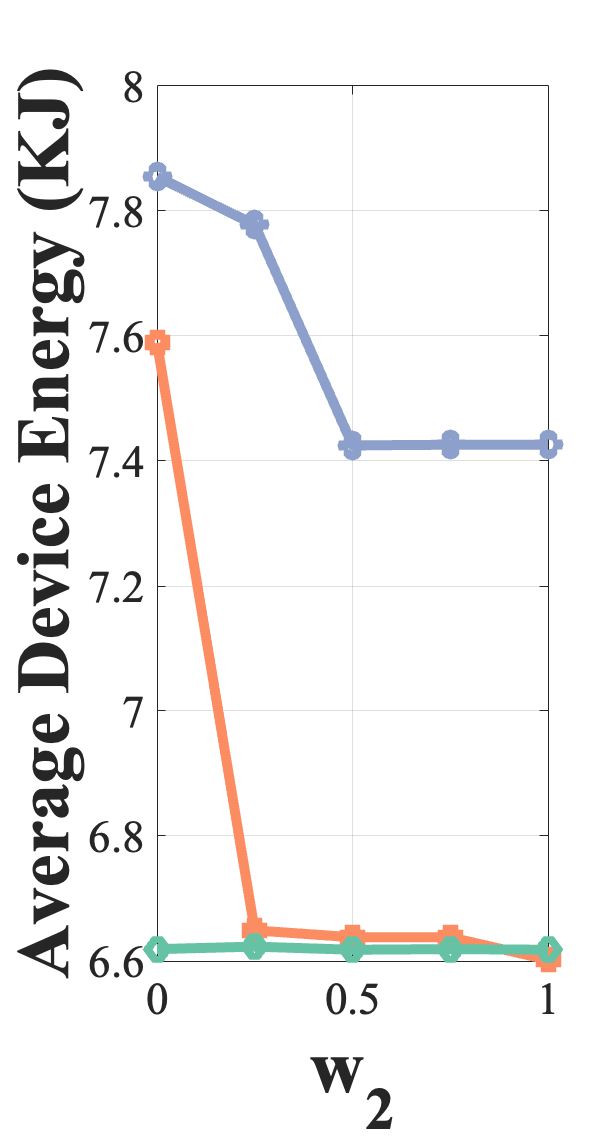}
        \label{fig:energy_latency}
    }\hspace{-0.1in}
    \subfigure[]{%
        \includegraphics[width=1.8cm]{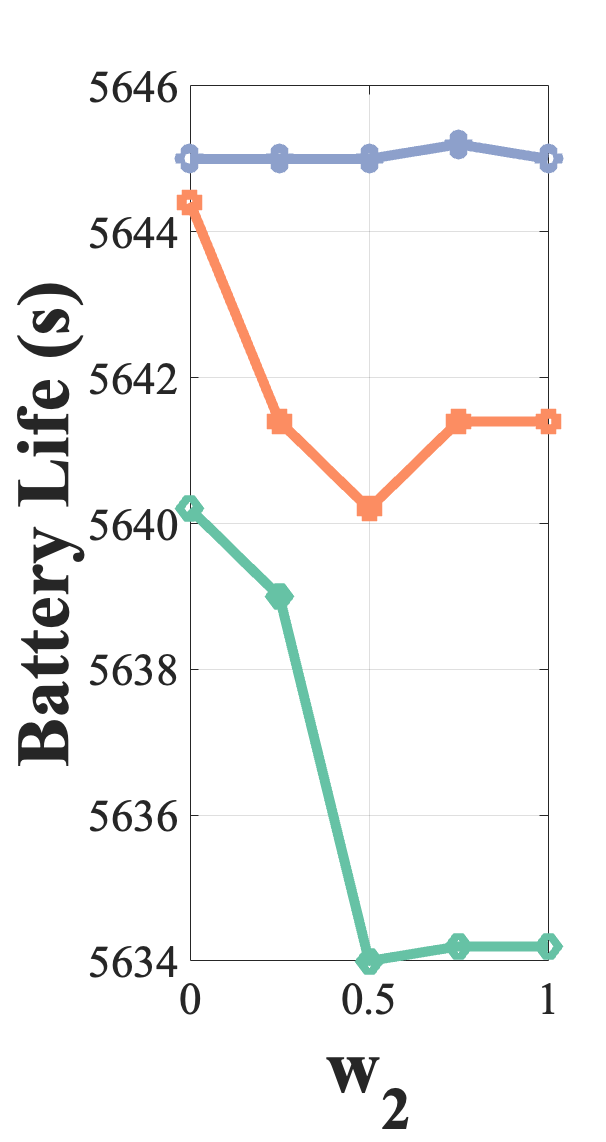}
        \label{fig:batterylife_latency}
    }\\
    \vspace{-0.15in}
    \subfigure{%
        \includegraphics[width=3cm]{Figures/EvaluationResults/ScreenShot03.png}
    }
    \vspace{-0.1in}
    \caption{System performance over varying latency weight}
    \vspace{-0.2in}
    \label{fig:latencyreward}
\end{figure}

Fig.~\ref{fig:latencyreward} presents the findings from similar experiments with latency reward weight manipulation. 
As we increase the emphasis on latency, there's a noticeable decrease in the average latency of the models. However, this reduction comes at the cost of increased inference energy consumption. Consequently, such a pattern inevitably leads to a diminished battery life for the devices. This observation is further corroborated when we compare Figs.~\ref{fig:latency_latency} and ~\ref{fig:latency_energy} where a clear tradeoff between latency and energy consumption emerges.
The selection of versions and cut points for the $w_2=0$ and $w_2=1$ models is detailed in Tab.~\ref{tab:versioncutpoint}. Notably, with a latency weight of $0$, the energy score predominance results in a greater proportion of layers being processed remotely. However, due to latency in transmission for $w_2=1$, offloading is postponed until later layers.

\begin{table}[b]
\vspace{-0.2in}
 \small
    \centering
    \caption{Cut point selection for reward weight manipulation}
    \vspace{-0.1in}
    \begin{tabular}{cccccc}
        \hline
        
        \multirow{2}{*}{Model} & \multirow{2}{*}{Version} & \multicolumn{4}{c}{Cut Point for} \\
        & & $w_2$ : 0&$w_2$ : 1 &$w_3$ : 0&$w_3$:1  \\
        \hline
        VGG & 19 & 5 & 10 & 10 & 5 \\
        ResNet & 50 & 3 & 13 & 13 & 3 \\
        DenseNet & 161 & 4 & 4 & 4 & 4 \\
        \hline
    \end{tabular}
    \label{tab:versioncutpoint}
    \vspace{-0.1in}
\end{table}

\subsubsection{Sensitivity of energy weight:}
\begin{figure}[t]
    \centering
    \subfigure[]{%
        \includegraphics[width=1.75cm]{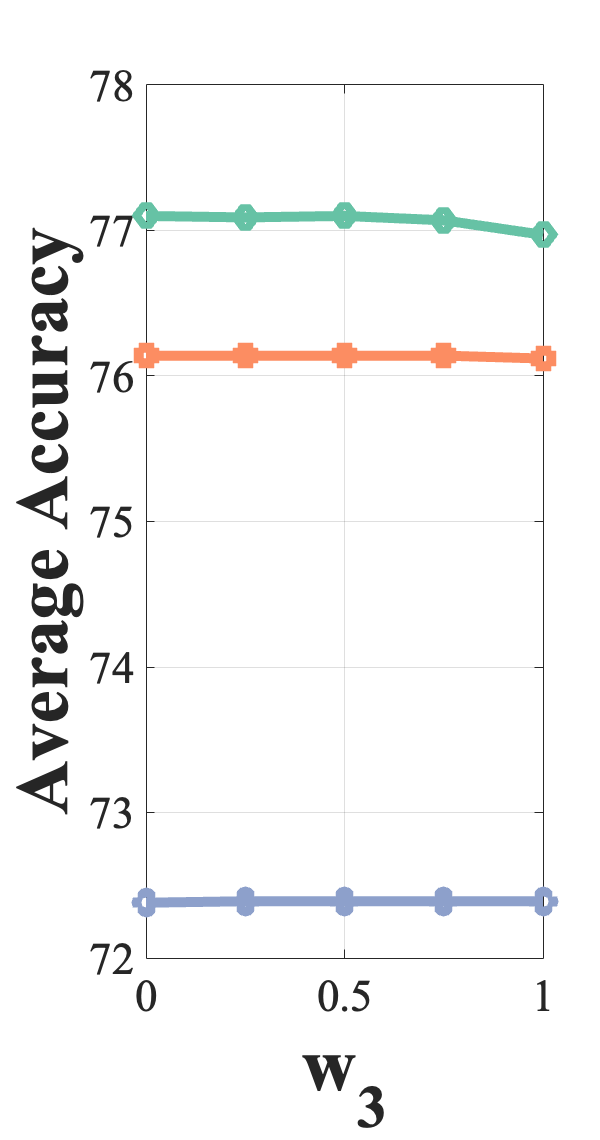}
        \label{fig:accuracy_energy}
    }\hspace{-0.1in}
    \subfigure[]{%
        \includegraphics[width=1.75cm]{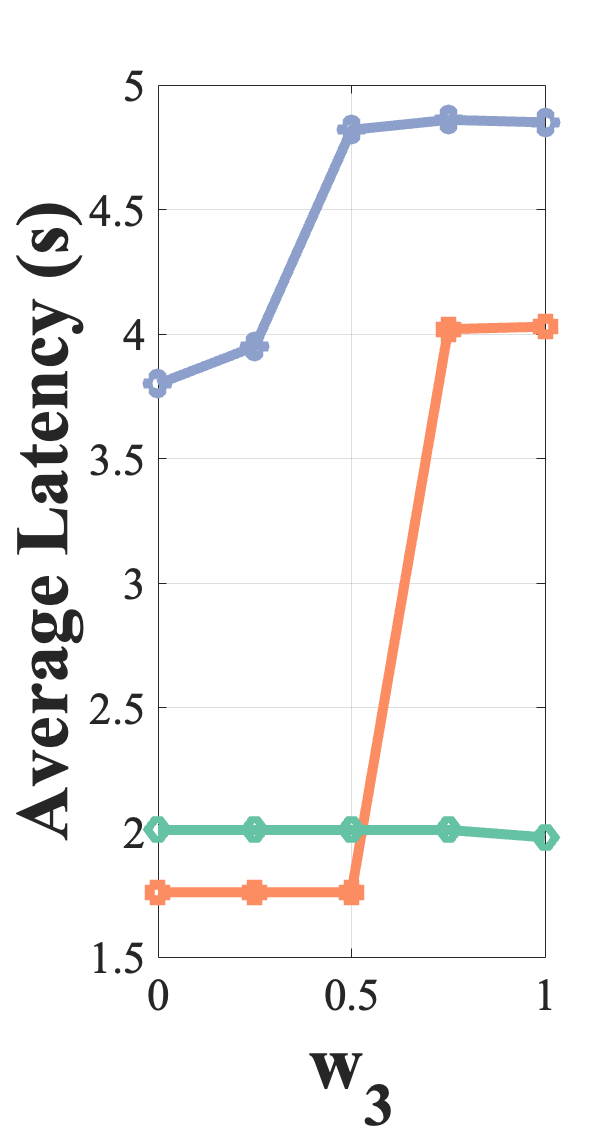}
        \label{fig:latency_energy}
    }\hspace{-0.1in}
    \subfigure[]{%
        \includegraphics[width=1.75cm]{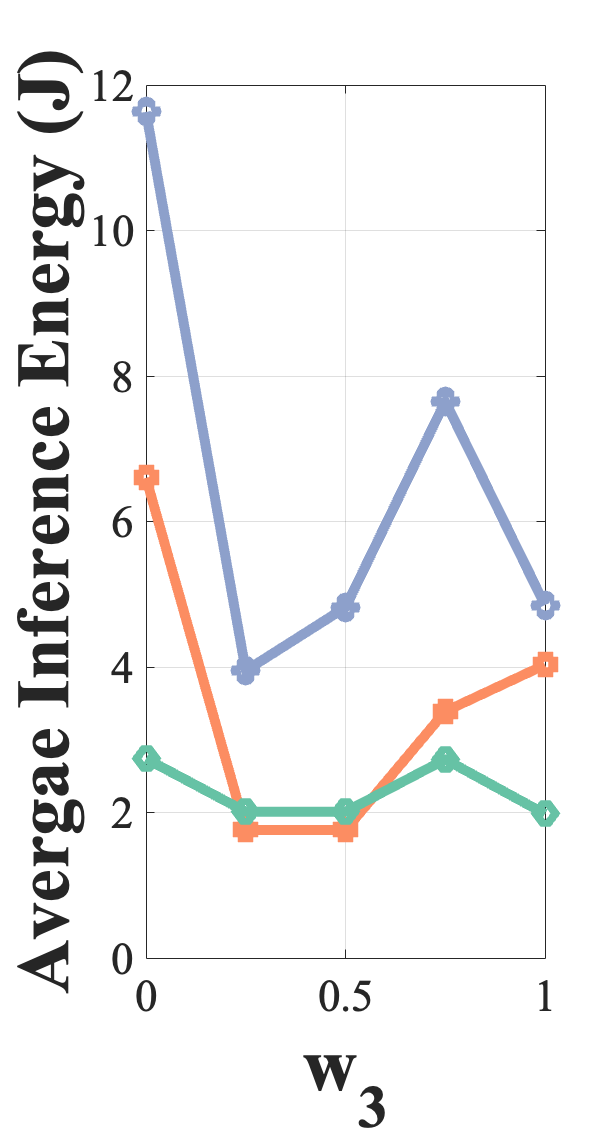}
        \label{fig:inference_energy}
    }\hspace{-0.1in}
    \subfigure[]{%
        \includegraphics[width=1.75cm]{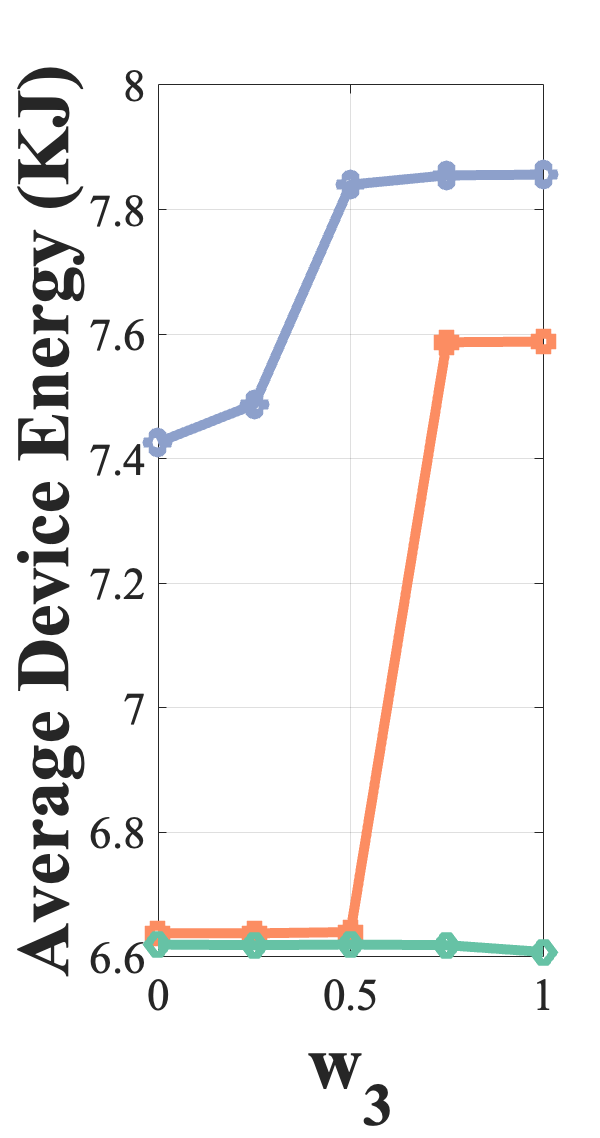}
        \label{fig:energy_energy}
    }\hspace{-0.1in}
    \subfigure[]{%
        \includegraphics[width=1.8cm]{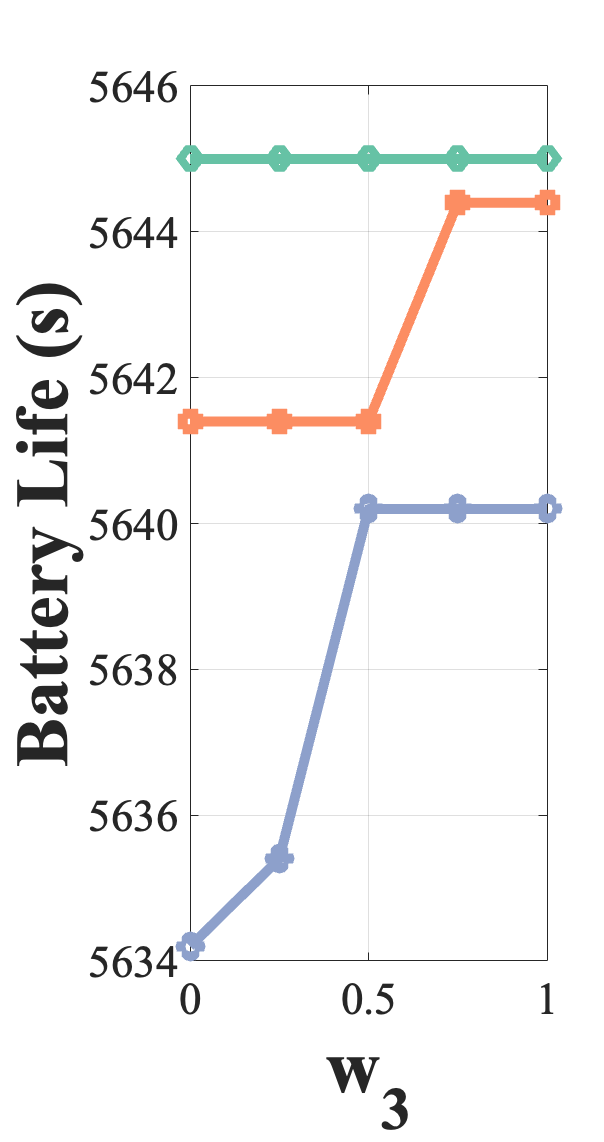}
        \label{fig:batterylife_energy}
    }\\
    \vspace{-0.15in}
    \subfigure{%
        \includegraphics[width=3cm]{Figures/EvaluationResults/ScreenShot03.png}
    }
    \vspace{-0.1in}
    \caption{System performance over varying energy weight}
    \vspace{-0.2in}
    \label{fig:energyreward}
\end{figure}

Next, Fig.~\ref{fig:energyreward} illustrates the findings from similar experiments with energy consumption reward weight manipulation. For obvious reasons with increasing the weight, the inference energy consumption drops as energy is given the higher priority. However, device/UAV energy consumption (from Fig.~\ref{fig:batterylife_energy}) shows a different trend, specifically, the device running \textit{DenseNet} - the most energy-consuming among the models. For accuracy performance, it can be observed that only \textit{ResNet} experiences a drop, as the energy savings in other models come from the choice of the optimal cut-point rather than the architecture (as seen in Tab.~\ref{tab:versioncutpoint}). 
It is also observed that when the energy reward is higher, the battery drains slower. E.g., the UAV running \textit{VGG} stays alive 6 more seconds when the model is energy efficient.
The choice of versions and cut points for these two models is also shown in Tab.~\ref{tab:versioncutpoint}. The interplay of energy and latency in cut point selection is evident.

\subsection{Effect of UAV Activity Profile}
\begin{figure}[htb]
    \centering
    \includegraphics[width = 7.5cm]{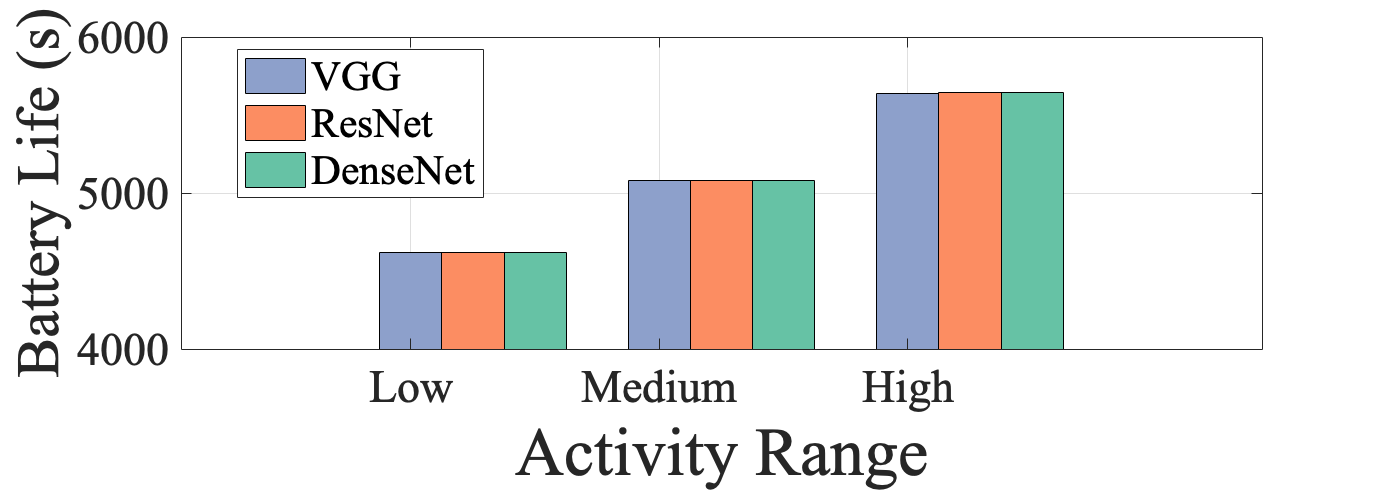}
    \vspace{-0.1in}
    \caption{UAV battery life for different DNNs}
    \label{fig:batterylife}
    \vspace{-0.2in}
\end{figure}

Finally, in Fig.~\ref{fig:batterylife}, we illustrate the impact of diverse UAV activity profiles on energy consumption by simulating 50 different activity profiles for each of the Low, Moderate, and High activity levels (from Tab.~\ref{tab:movement_distribution}). All the experiments are conducted for high data rate, i.e., for WiFi connectivity. 
We observe that DNN model energy consumption itself is minimal compared to the energy expenditure during kinetic activities. Consequently, we do not observe significant differences between DNN models. However, it is noteworthy that vertical movement incurs the highest energy consumption. Thus, increasing its probability directly correlates to faster drainage, explaining quicker drainage for Low Activity.

\section{Conclusions}
\label{sec:conclusion}
In this paper, we analyzed the end-to-end latency vs. inference accuracy vs. device energy consumption trade-off for `just-in-time' edge-AI implementations  
and proposed {\em Infer-EDGE} framework that employs a novel A2C based RL model. 
The {\em Infer-EDGE} framework performs DNN version selection and cut point selection based on resource availability and system performance requirements. 
We demonstrated how the underlying A2C based RL agent learnt about the environment through actions and rewards which eventually converged at an optimal trade-off point for involved performance metrics. 
Using real world DNNs and a hardware testbed, we evaluated the benefits of {\em Infer-EDGE} in terms of device energy saving, accuracy improvement, and end-to-end latency reduction. 

\bibliographystyle{IEEEtran}
\bibliography{reference}

\end{document}